\documentclass[11pt,a4paper]{article}

\usepackage[margin=2.2cm]{geometry}
\usepackage{amsmath,amssymb}
\usepackage{graphicx}
\usepackage{booktabs}
\usepackage{multirow}
\usepackage{xcolor}
\usepackage[numbers,sort&compress]{natbib}
\usepackage[colorlinks=true,linkcolor=blue,citecolor=blue,urlcolor=blue]{hyperref}
\usepackage{caption}
\usepackage{subcaption}
\usepackage{authblk}
\newcommand{\mH}{m_H}
\newcommand{\cba}{c_{\beta-\alpha}}
\newcommand{\tb}{t_\beta}
\newcommand{\CLs}{\mathrm{CL}_s}

\newcommand{\Ztwod}{Z_{\rm 2D}}
\newcommand{\sbr}{\sigma\!\times\!\mathrm{BR}}
\newcommand{\mll}{m_{\ell\ell}}

\title{
	Machine-learning discovery potential for a heavy Higgs boson \\
	in $e^+e^-\to ZH$ at a future lepton collider
}
\author{Majid Hashemi\thanks{Email: \texttt{hashemi\_mj@shirazu.ac.ir}}}
\affil{Department of Physics, College of Science, Shiraz University, Shiraz, 71946-84795, Iran}
\date{\today}

\begin{document}
	\maketitle
	
	% =====================================================================
\begin{abstract}
	A study of the discovery potential for a CP-even heavy neutral Higgs
	boson $H$ is presented in the Type-I two-Higgs-doublet model (2HDM) at a future
	$e^+e^-$ collider with $\sqrt{s}=500~\mathrm{GeV}$, focusing on the
	process $e^+e^-\to ZH$ with $H\to WW$, and with $Z\to\ell\ell$ or
	$Z\to\nu\nu$.  The final states are four jets plus two leptons
	($4j2\ell$) and four jets plus missing transverse momentum
	($4j2\nu$).  Both a Boosted Decision Tree (BDT) and a Deep Neural Network (DNN) were 
trained on the same inputs. The BDT was found to be substantially faster in training and inference 
without any loss of performance, and it is therefore the classifier used in the interpretation.
Three strategies are compared: a cut-based analysis, a one-dimensional BDT selection trained per mass point, and a two-dimensional (BDT $\times$ $\mH$) template
	analysed with a profile likelihood.  The 2D template improves the
	expected discovery significance by a factor of $2.5$--$3.6$ in the
	$4j2\ell$ channel and by a factor of $5$--$18$ in the $4j2\nu$ channel
	relative to the cut-based analysis, with the largest gains at high
	$\mH$ where the cut-based mass window loses efficiency.  We translate
	the expected exclusions into $95\%$ CL upper limits on $\sbr$ and
	discuss the implications for the Type-I 2HDM parameter space.
\end{abstract}
	
	% =====================================================================
	\section{Introduction}
	\label{sec:intro}
	
	The discovery of a Higgs boson with mass $125~\mathrm{GeV}$ by ATLAS and
	CMS~\cite{HiggsATLAS,HiggsCMS} established the mechanism of electroweak
	symmetry breaking, but the structure of the scalar sector remains an open
	question.  Extended scalar sectors, and in particular the
	two-Higgs-doublet model (2HDM)~\cite{2hdm1,2hdm2,2hdm3,2hdm_TheoryPheno,hhg},
	predict additional neutral and charged scalars whose discovery would
	directly probe the shape of the scalar potential.  In the Type-I
	realisation, all fermions couple to a single doublet, so the Yukawa
	couplings are controlled by a common factor $\cot\beta$ and the
	phenomenology is governed by $\tan\beta$ and the alignment parameter
	$\cba$.  The current LHC constraints on heavy neutral
	scalars~\cite{ATLASNeutralHiggs,CMSNeutralHiggs,atlas1,atlas2,cms1,cms2,cms3}
	and on the observed $125~\mathrm{GeV}$ state~\cite{LHC1,CMS22,portrait}
	leave substantial room for a heavy CP-even $H$ in the intermediate-mass
	region, and motivate a dedicated study at a future lepton collider.

Direct LHC searches in bosonic channels provide complementary constraints on the mis-aligned region. A search for a heavy Higgs decaying to $WW$ excludes SM-like couplings for masses up to about $1870~\mathrm{GeV}$ at $95\%$ CL~\cite{CMSWW2020}, and in the Type-I 2HDM ATLAS excludes resonance masses in the range $270$--$810~\mathrm{GeV}$ for $\tan\beta=10$ and $\cos(\beta-\alpha)=-0.1$ in the $X\to HH$ channel~\cite{ATLAS2HDMHH}. The region relevant to this work ($|\cba|\sim 0.1$, $\tan\beta\lesssim 10$, $\mH\lesssim 240~\mathrm{GeV}$) remains weakly constrained by these searches, because the $H\to WW$ and $H\to hh$ branching ratios are suppressed near the LHC kinematic limit~\cite{Hashemi2026}.

	The dominant decay modes of a heavy CP-even $H$ in the Type-I 2HDM
	depend strongly on $\tan\beta$: at moderate $\tan\beta$ the
	$WW$ and $hh$ channels are competitive with $b\bar b$ and
	$\tau\tau$, and they offer clean multi-lepton or multi-jet signatures
	that are complementary to the standard fermionic searches.  In this
	paper we focus on the process $e^+e^-\to ZH$, with the decay chains
	\begin{align}
		H \to WW \to 4j, \quad Z \to 2\ell \quad &(4j2\ell), \\
		H \to WW \to 4j, \quad Z \to 2\nu \quad &(4j2\nu),
	\end{align}. The $4j2\ell$ and
	$4j2\nu$ final states are characterised by a four-jet system recoiling
	against a $Z$ boson, with the $H$ mass reconstructed from the
	four-jet system.  The $ZH$ cross section is evaluated at
	$\sqrt{s}=500~\mathrm{GeV}$, and the $Z$ boson is reconstructed either
	from the two leptons ($4j2\ell$) or from the missing four-momentum
	($4j2\nu$); the recoil mass against the $Z$ candidate provides an
	independent handle on $\mH$.
	
	The recoil-mass technique, originally proposed for lepton colliders~\cite{Bjorken1977}, has since become a benchmark for model-independent $ZH$ cross-section measurements at future Higgs factories~\cite{Thomson2016,Marshall2016}. Recent FCC-ee studies have demonstrated that the recoil mass against a
	leptonically decaying $Z$ provides a clean handle on the inclusive
	$e^+e^-\to HZ$ cross section and on $|g_{HZZ}|$, with a Higgs mass
	resolution of order $0.25~\mathrm{GeV}$ for the $125~\mathrm{GeV}$ state
	at $\sqrt{s}=240~\mathrm{GeV}$~\cite{FCCeeRecoil2025}. The present
	analysis is complementary to this programme: rather than using the recoil
	mass to measure the inclusive $ZH$ rate, we reconstruct the heavy $H$
	directly from its $WW$ decay products, which separates the
	mis-aligned Type-I 2HDM signal from the $t\bar{t}$, $WWZ$, and $ZZZ$
	backgrounds using the decay kinematics, and is therefore sensitive to
	the $H\to WW$ and $H\to hh$ branching ratios in addition to the
	production coupling.
	
	Analyses of this kind have traditionally relied on sequential cuts on
	kinematic variables~\cite{HG_2,MH_1,HG_4}.
	Machine-learning classifiers can exploit correlations between these
	variables that are invisible to a cut-based selection.  The goal of
	this paper is to quantify that gain in a controlled way, using the same
	event sample, the same object definition, and the same statistical
	machinery, so that the comparison isolates the effect of the classifier
	itself.
	
	Our main results are:
	\begin{itemize}
		\item a one-dimensional BDT scan per mass point, which already
		improves the expected significance by a factor of
		$1.7$--$1.9$ ($4j2\ell$) and $3.4$--$8.9$ ($4j2\nu$) over the
		cut-based analysis;
		\item a two-dimensional $(\mathrm{BDT},\mH)$ template analysed with a
		profile likelihood, which further improves the significance by
		a factor of $2.5$--$3.6$ ($4j2\ell$) and $5.3$--$18.3$
		($4j2\nu$);
		\item $95\%$ CL upper limits on $\sbr$ as a function of $\mH$, and
		the corresponding exclusion contours in the
		$(\cba,\tb)$ plane.
	\end{itemize}
	
This work extends the cut-based analysis of
Ref.~\cite{Hashemi2026}, which studied the same Type-I 2HDM
parameter space in the $e^+e^-\to ZH$ channel at a future
$e^+e^-$ collider.  The event sample, the object definitions, and the
selection strategy are the same; the difference is the statistical
treatment of the classifier output.  We therefore compare the
cut-based analysis of Ref.~\cite{Hashemi2026} with a 1D BDT and a
2D $(\mathrm{BDT},\mH)$ template on the same events, so that the
comparison isolates the effect of the classifier itself. Two further differences in the statistical treatment should be borne in
mind: the present analysis uses the profiled Asimov significance
introduced in Sec.~\ref{sec:stats} rather than the simple Asimov
formula adopted in Ref.~\cite{Hashemi2026}, and it includes a
jet energy scale uncertainty together with a $10\%$
background normalisation uncertainty in the profile, which yields a
more realistic significance than the previously published result.
	
	The paper is organised as follows.  Section~\ref{sec:theory} recalls the
	relevant features of the Type-I 2HDM.  Section~\ref{sec:generation}
	describes the event generation and object reconstruction.
	Section~\ref{sec:selection} details the event selection and the feature
	set.  Section~\ref{sec:ml} presents the machine-learning classifiers.
	Section~\ref{sec:stats} describes the statistical methodology.
	Section~\ref{sec:results} presents the results, and
	Section~\ref{sec:conclusions} concludes.
	
	% =====================================================================
	\section{The Type-I two-Higgs-doublet model}
	\label{sec:theory}
	
The 2HDM extends the Standard Model by a second scalar doublet.  We
consider the CP-conserving Type-I realisation, in which all quarks and
leptons couple to the same doublet, so that the Yukawa couplings are
controlled by a common factor $\cot\beta$.  After electroweak symmetry
breaking the two doublets acquire vacuum expectation values
$v_1,v_2$ with $v_1^2+v_2^2=v^2=(246~\mathrm{GeV})^2$ and
$\tan\beta\equiv v_2/v_1$.  The Yukawa Lagrangian reads
\begin{equation}
	\mathcal{L}_Y = -\,\frac{1}{\sqrt{2}}\,
	\bar Q_L \left( \frac{\cos\alpha}{\sin\beta}\,h
	+ \frac{\sin\alpha}{\sin\beta}\,H
	- i\,\frac{\cot\beta}{\sqrt{2}}\,\gamma_5 A
	\right) Q_L + \text{h.c.},
	\label{eq:yukawa}
\end{equation}
where $\alpha$ is the mixing angle in the CP-even sector and $h$ is
identified with the observed $125~\mathrm{GeV}$ state.

The scalar potential of the CP-conserving 2HDM with a softly broken
$Z_2$ symmetry (under which $\Phi_1\to\Phi_1$, $\Phi_2\to-\Phi_2$) is
\begin{equation}
	\begin{split}
		V(\Phi_1,\Phi_2) = {} & m_{11}^2\,\Phi_1^\dagger\Phi_1
		+ m_{22}^2\,\Phi_2^\dagger\Phi_2
		- m_{12}^2\!\left(\Phi_1^\dagger\Phi_2 + \mathrm{h.c.}\right) \\
		& + \frac{\lambda_1}{2}(\Phi_1^\dagger\Phi_1)^2
		+ \frac{\lambda_2}{2}(\Phi_2^\dagger\Phi_2)^2
		+ \lambda_3(\Phi_1^\dagger\Phi_1)(\Phi_2^\dagger\Phi_2) \\
		& + \lambda_4(\Phi_1^\dagger\Phi_2)(\Phi_2^\dagger\Phi_1)
		+ \frac{\lambda_5}{2}\!\left[(\Phi_1^\dagger\Phi_2)^2
		+ \mathrm{h.c.}\right],
	\end{split}
	\label{eq:potential}
\end{equation}
where the $m_{12}^2$ term is the only soft-breaking term.  The
physical CP-even mass eigenstates are obtained by rotating the two
doublets by the angle $\alpha$,
\begin{equation}
	h = -\Phi_1^0\sin\alpha + \Phi_2^0\cos\alpha, \qquad
	H = \phantom{-}\Phi_1^0\cos\alpha + \Phi_2^0\sin\alpha,
	\label{eq:rotation}
\end{equation}
so that the alignment limit $\cba\to 0$ corresponds to $h$ being
SM-like.

The couplings of the two CP-even states to gauge bosons and fermions,
normalised to the corresponding SM Higgs couplings, are collected in
Table~\ref{tab:couplings}.  In the alignment limit $h$ has SM-like
couplings to all gauge bosons and fermions, while the heavier $H$
decouples from the gauge sector; at intermediate values of $\cba$ the
$HWW$ and $HZZ$ couplings are proportional to $\cos(\beta-\alpha)$
and drive the $WW$ and $ZZ$ partial widths.  In the Type-I
realisation, the up- and down-type fermions share the same
$\cos\alpha/\sin\beta$ and $\sin\alpha/\sin\beta$ factors, which is
what distinguishes Type-I from Type-II and what makes the fermionic
decays of $H$ suppressed at large $\tan\beta$.

\begin{table}[t]
	\centering
	\caption{Couplings of the CP-even states $h$ and $H$ to gauge
		bosons and fermions in the Type-I 2HDM, normalised to the SM
		Higgs couplings.  In the alignment limit $\cba\to 0$ the
		column for $h$ reduces to unity and the column for $H$ reduces
		to zero, so that the heavier state decouples from the gauge
		sector.}
	\label{tab:couplings}
	\begin{tabular}{lcc}
		\toprule
		Coupling & $h$ (SM-like) & $H$ (heavy) \\
		\midrule
		$VV$ ($V=W,Z$)                       & $\sin(\beta-\alpha)$ & $\cos(\beta-\alpha)$ \\
		$t\bar t$, $b\bar b$, $\tau^+\tau^-$ & $\cos\alpha/\sin\beta$ & $\sin\alpha/\sin\beta$ \\
		\bottomrule
	\end{tabular}
\end{table}

We work with the \textsc{2HDMC} convention~\cite{2hdmc1,2hdmc2,2hdmc3}, in which the
free parameters are $\mH$, $m_A$, $m_{H^\pm}$, $\tan\beta$, $\cba$,
and the soft $Z_2$-breaking mass $m_{12}^2$.  For the phenomenological
study in this paper we fix $m_A=m_{H^\pm}=\mH$ and $\mH$ in the range
$160$--$240~\mathrm{GeV}$, and scan over $(\cba,\tb)$.

The partial widths $\Gamma(H\to WW,ZZ,t\bar t,b\bar b,\tau^+\tau^-)$
are computed with \textsc{2HDMC}, which includes the tree-level gauge
and scalar channels and the one-loop QCD corrections to the fermionic
channels; the total width and the branching ratios are obtained by
summing all open channels.  At moderate $\tan\beta$ the $WW$ channel is competitive with $b\bar b$, which is the region targeted
by the $4j2\ell$ and $4j2\nu$ searches studied in this paper.  At
large $\tan\beta$ the fermionic channels dominate, the $H\to WW$ branching ratio decreases, and the sensitivity of the
present search drops accordingly; this is reflected in the exclusion
contours of Sec.~\ref{sec:contours}.
	
	% =====================================================================
	\section{Event generation and object reconstruction}
	\label{sec:generation}
	
Signal and background events are generated at leading order with
\textsc{WHIZARD}~\cite{whizard1,whizard2} using the Type-I 2HDM
implementation.  The beam-energy spectrum is described by
\textsc{Circe2}~\cite{circe2}, which accounts for the effects of
initial-state radiation and beamstrahlung at the nominal
$\sqrt{s}=500~\mathrm{GeV}$; only the $e^+e^-$ luminosity is used, and
photon-induced processes are not generated separately.  Parton showers
are generated with \textsc{Pythia}\,8~\cite{pythia}, and the detector
response is simulated with \textsc{Delphes}\,3~\cite{delphes1,delphes2,delphes3}
using the ILCgen detector card.  Jets are clustered with the anti-$k_t$
algorithm~\cite{antikt} with radius $R=0.5$; leptons are required to be
isolated.  The $Z\to\ell\ell$ and $Z\to\nu\nu$ hypotheses are selected
by requiring either two isolated leptons or large missing
four-momentum, respectively.
 The centre-of-mass energy is $\sqrt{s}=500~\mathrm{GeV}$.  In the
	$4j2\ell$ channel the recoil mass is computed from the two leptons as
	\begin{equation}
		m_{\rm recoil}^2 = \bigl(\sqrt{s} - (E_{\ell^+} + E_{\ell^-})\bigr)^2
		- \bigl|\vec p_{\ell^+} + \vec p_{\ell^-}\bigr|^2,
		\label{eq:mrecoil-ll}
	\end{equation}
	where $E_{\ell^\pm}$ and $\vec p_{\ell^\pm}$ are the energies and
	three-momenta of the two leptons; no mass constraint is imposed on the
	$\ell\ell$ system.  In the $4j2\nu$ channel the same formula is
	applied to the reconstructed $H$ candidate,
	$m_{\rm recoil}^2 = (\sqrt{s}-E_H)^2 - |\vec p_H|^2$, but the
	resulting variable is not used as a classifier input
	(Sec.~\ref{sec:selection}).
	
	The relevant backgrounds at $\sqrt{s}=500~\mathrm{GeV}$ are
	$e^+e^-\to t\bar t$, $e^+e^-\to t\bar t b\bar b$,
	$e^+e^-\to WW$, $e^+e^-\to ZZ$, $e^+e^-\to WWZ$, and
	$e^+e^-\to ZZZ$.  Background events are generated with the same chain
	and normalised to their cross sections at $\sqrt{s}=500~\mathrm{GeV}$.
	The signal cross section is evaluated at the reference point
	$(\cba,\tb)=(0.2,5)$ and rescaled by the ratio of
	$(\sigma\cdot\mathrm{BR})$ at each grid point, so that all events share
	the same kinematic shape.  The selection strategy --- the object
	definitions, the $W$-pairing algorithm, and the final shape cuts on
	$\Delta R_{WW}$, $\beta_H$, and $\cos\theta_H$ --- is the same as in
	the cut-based analysis of Ref.~\cite{Hashemi2026}.  The cut-based
	significances quoted in Sec.~\ref{sec:results} are obtained by applying
	this selection to the same event sample used to train the classifiers,
	so that the ML-vs-cut comparison isolates the effect of the classifier
	alone.
	
The detector response is modelled with \textsc{Delphes}\,3~\cite{delphes1,delphes2,delphes3}
using the ILCgen card~\cite{ILCgen}, a generic ILC detector model
based on the ILD detector concept.

Jets are reconstructed with the anti-$k_t$ algorithm~\cite{antikt}
($R=0.5$, $p_T>5~\mathrm{GeV}$) and smeared with a resolution of
$0.3\%$ for $|\eta|<0.76$ and $1.5\%$ for $|\eta|\ge0.76$, following
the ILD performance parametrisation; the $\pm1\sigma$ variation of
this resolution is the JES uncertainty used in Sec.~\ref{sec:jes}.
Jets are $b$-tagged at the 50\% working point, which is the working
point used in the event selection of Sec.~\ref{sec:selection}.

The ILCgen card does not include a jet energy smearing to mimic the
$\gamma\gamma\to\mathrm{hadrons}$ overlay background, whose residual
effect is treated as part of the JES systematic uncertainty discussed
in Sec.~\ref{sec:jes}.
	
	The full simulation and analysis chain is: \textsc{WHIZARD}~\cite{whizard1,whizard2}
	for event generation, \textsc{Pythia}\,8~\cite{pythia} for parton
	showering, \textsc{Delphes}\,3~\cite{delphes1,delphes2,delphes3} with
	the ILCgen card for detector simulation, \textsc{scikit-learn}~\cite{sklearn}
	and XGBoost~\cite{xgboost} for the BDT, \textsc{TensorFlow/Keras}~\cite{tensorflow}
	for the DNN, \textsc{shap}~\cite{shap} for feature importance,
	\textsc{pyhf}~\cite{pyhf} for the profile-likelihood and $\CLs$
	calculations, and \textsc{matplotlib}~\cite{matplotlib} and
	\textsc{numpy}~\cite{numpy} for plotting and numerics.
	% =====================================================================
	\section{Event selection and feature set}
	\label{sec:selection}
	
	Events are required to contain at least four jets, and either two
	isolated leptons ($4j2\ell$ channel) or large missing four-momentum
	($4j2\nu$ channel).  The four leading jets are used to reconstruct the
	heavy-Higgs candidate; the $Z$ candidate is reconstructed from the two
	leptons ($4j2\ell$) or from the missing four-momentum ($4j2\nu$).  A
	kinematic fit is applied to improve the $\mH$
	resolution~\cite{ALEPH:4CKinematicFit}.  The same preselection is
	applied to signal and background, and follows the cut-based strategy of
	Ref.~\cite{Hashemi2026}, re-implemented for this analysis.  Because
	the cut-based and ML analyses use the same events, the same object
	definitions, and the same statistical machinery, the comparison in
	Sec.~\ref{sec:results} isolates the gain from the classifier itself.
	
	For each event we construct the feature vector used by the
	classifier.  In the $4j2\ell$ channel the features are
	\begin{multline}
		\{\mll, E_T^{\rm miss}, m_{\rm recoil}, m_{W_1}, m_{W_2}, \mH,
		\Delta R_{WW}, \beta_H, p_H, \cos\theta_H, \\
		p_T^{\ell_1}, p_T^{\ell_2}, \eta_{\ell_1}, \eta_{\ell_2},
		p_T^{W_1}, p_T^{W_2}, \eta_{W_1}, \eta_{W_2}, \\
		p_T^{j_1},\dots,p_T^{j_4}, \eta_{j_1},\dots,\eta_{j_4},
		\Delta R_{12},\dots,\Delta R_{34}, N_{\rm jets}, N_{\rm leptons}\}.
	\end{multline}
	The $4j2\nu$ channel uses the same set with two differences: the
	lepton-specific variables ($\mll$, $p_T^{\ell_1}$, $p_T^{\ell_2}$,
	$\eta_{\ell_1}$, $\eta_{\ell_2}$) are absent because there are no
	reconstructed leptons, and $m_{\rm recoil}$ is not used, because in
	this channel it is defined from the $H$ candidate and was found not to
	improve the separation beyond the other kinematic variables.  The
	choice of features is motivated by the kinematic reconstruction of the
	$H\to WW$ decay, and by the previous cut-based
	analysis~\cite{Hashemi2026}.
	
	% =====================================================================
	\section{Machine-learning classifiers}
	\label{sec:ml}
\subsection{Boosted decision tree}
\label{sec:xgb}

We train a gradient-boosted decision tree (XGBoost~\cite{xgboost})
separately for each mass point and each channel.  The training set is
formed by merging the signal sample at that mass with the full
background sample, with weights equal to the expected event yields at
$\mathcal{L}=10~\mathrm{fb}^{-1}$.  The data are split into $70\%$
training and $30\%$ testing, stratified by class.  We use a standard
XGBoost configuration designed to limit overtraining on this dataset:
300 trees with a maximum depth of 5, a learning rate of 0.1, and
column and row subsampling at 80\%.  No hyperparameter optimisation
was performed.

\begin{figure}[t]
	\centering
	\includegraphics[width=0.48\textwidth]{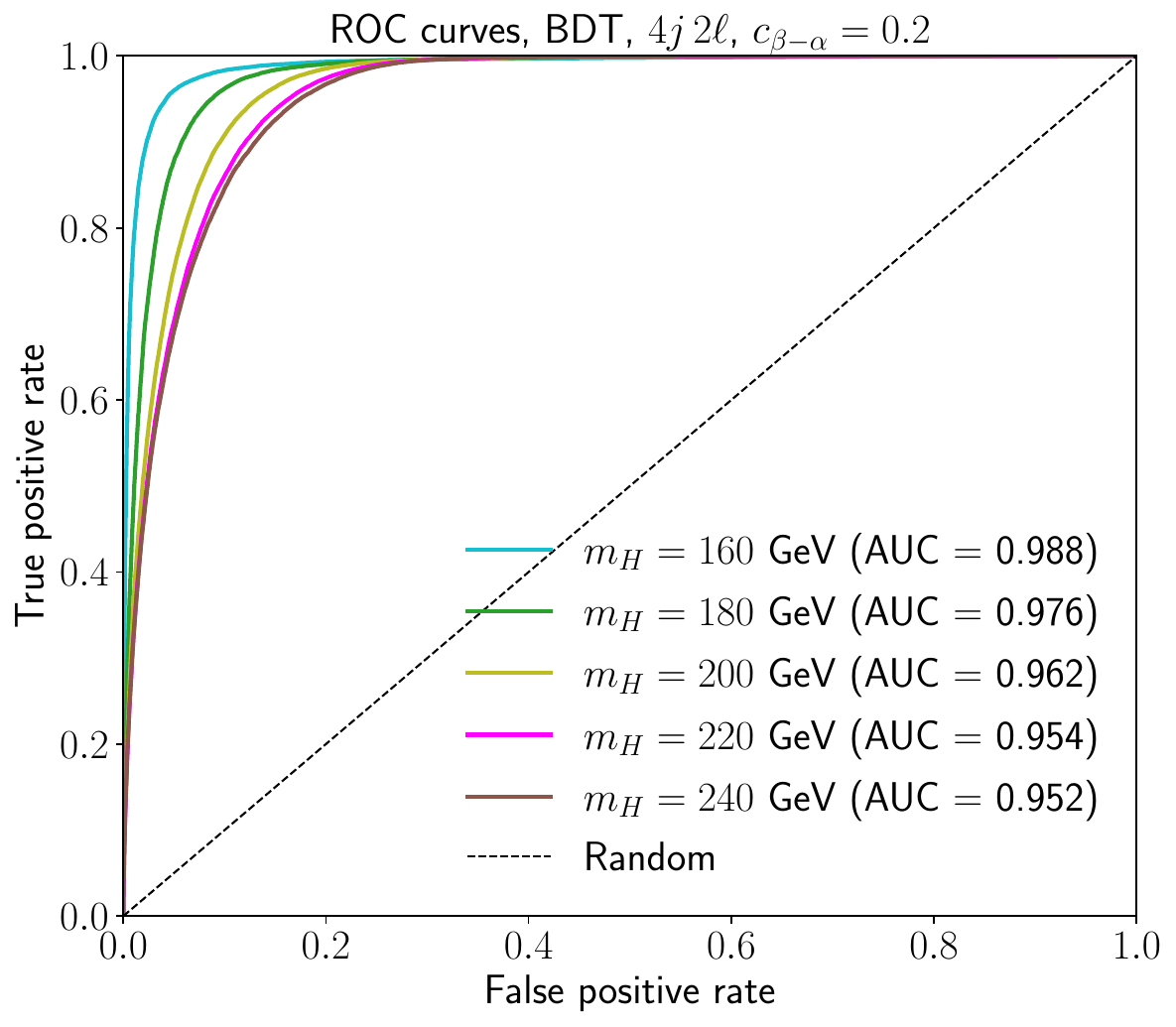}
	\includegraphics[width=0.48\textwidth]{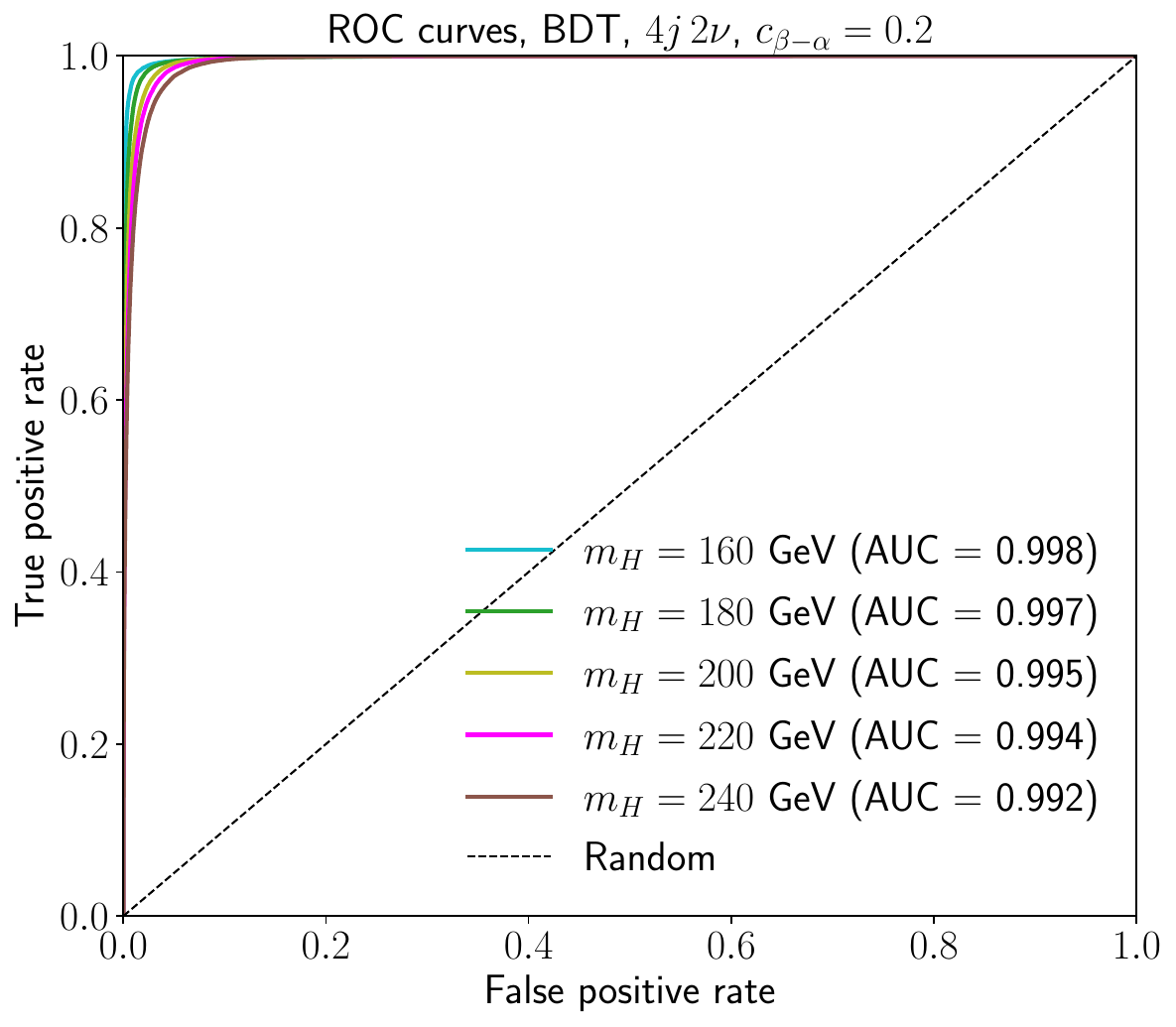}
	\caption{ROC curves for the XGBoost classifier in the $4j2\ell$
		(left) and $4j2\nu$ (right) channels, for all mass points.  The
		AUC values quoted in the legend are weighted by the expected event
		yields.}
	\label{fig:roc}
\end{figure}

Figure~\ref{fig:roc} shows the ROC curves for all mass points in the
two channels.  The weighted AUC decreases from $0.988$ at
$\mH=160~\mathrm{GeV}$ to $0.951$ at $\mH=240~\mathrm{GeV}$ in the
$4j2\ell$ channel, and from $0.998$ to $0.992$ in the $4j2\nu$
channel.  The degradation with mass reflects the reduced $S/B$ and
the broader $\mH$ resolution at higher mass, which are precisely the
regions where the 2D template provides the largest gain (see
Sec.~\ref{sec:z2d}).

\begin{figure}[t]
	\centering
	\includegraphics[width=\textwidth]{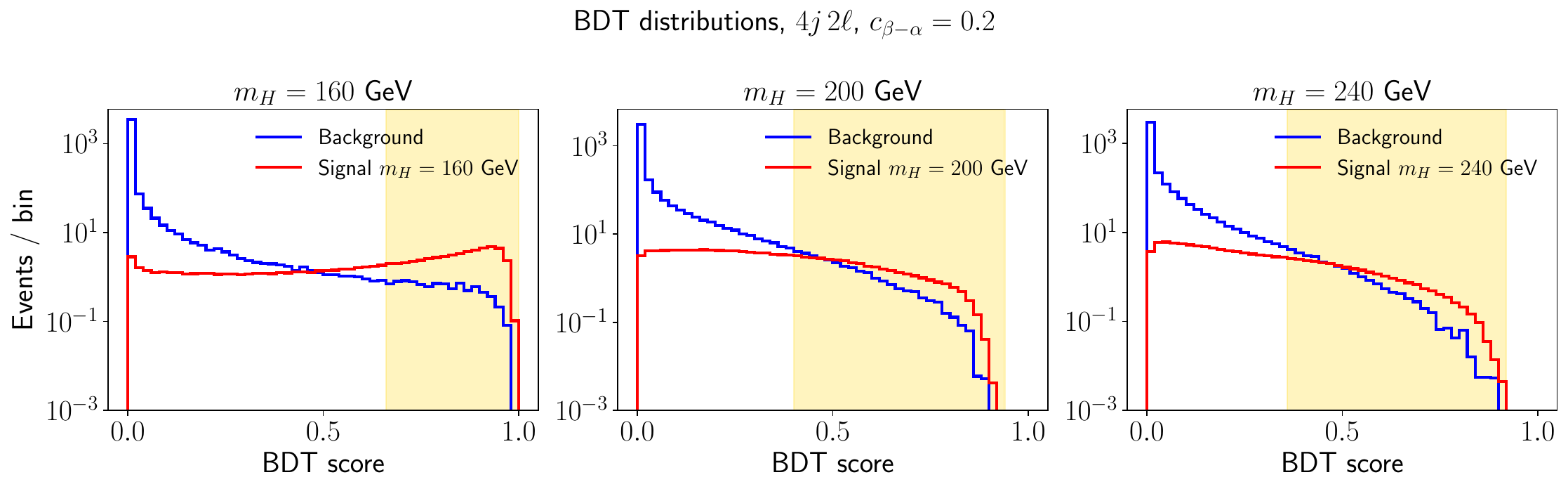}\\
	\includegraphics[width=\textwidth]{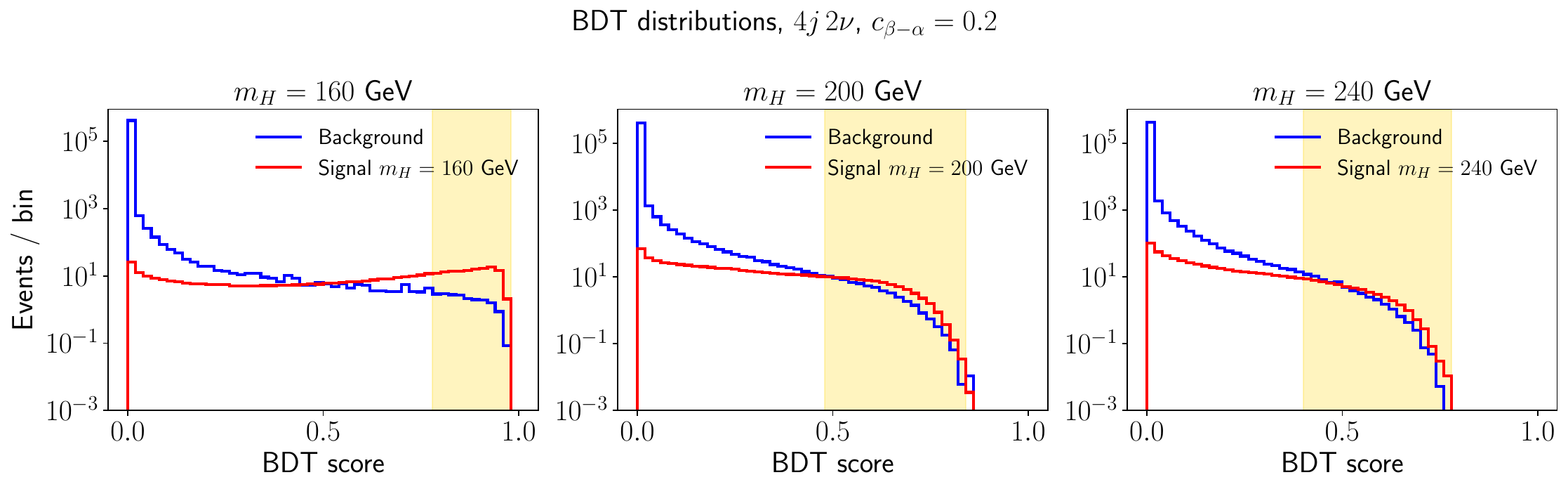}
	\caption{BDT score distributions for background (blue) and signal
		(red) at $\mH=160,200,240~\mathrm{GeV}$, for the $4j2\ell$ (top)
		and $4j2\nu$ (bottom) channels.  The gold band indicates the
		optimal 1D BDT window, i.e.\ the window that maximises the
		profiled Asimov significance of Eq.~(\ref{eq:asimov}).  The
		$y$-axis is logarithmic, so the high-score tail where the signal
		sits is visible.}
	\label{fig:bdt_dist}
\end{figure}

Figure~\ref{fig:bdt_dist} shows the BDT score distributions for the
total background and for the signal at the three representative
masses, with the optimal 1D BDT window shaded in gold.  The
distributions are shown on a logarithmic scale so that the high-score
tail --- where the signal sits --- is visible; on a linear scale the
tail would be buried under the much larger background at low score.
The separation between the red (signal) and blue (background) curves
increases with $\mH$, which is the same trend visible in the ROC
curves (Fig.~\ref{fig:roc}).

\begin{figure}[t]
	\centering
	\includegraphics[width=\textwidth]{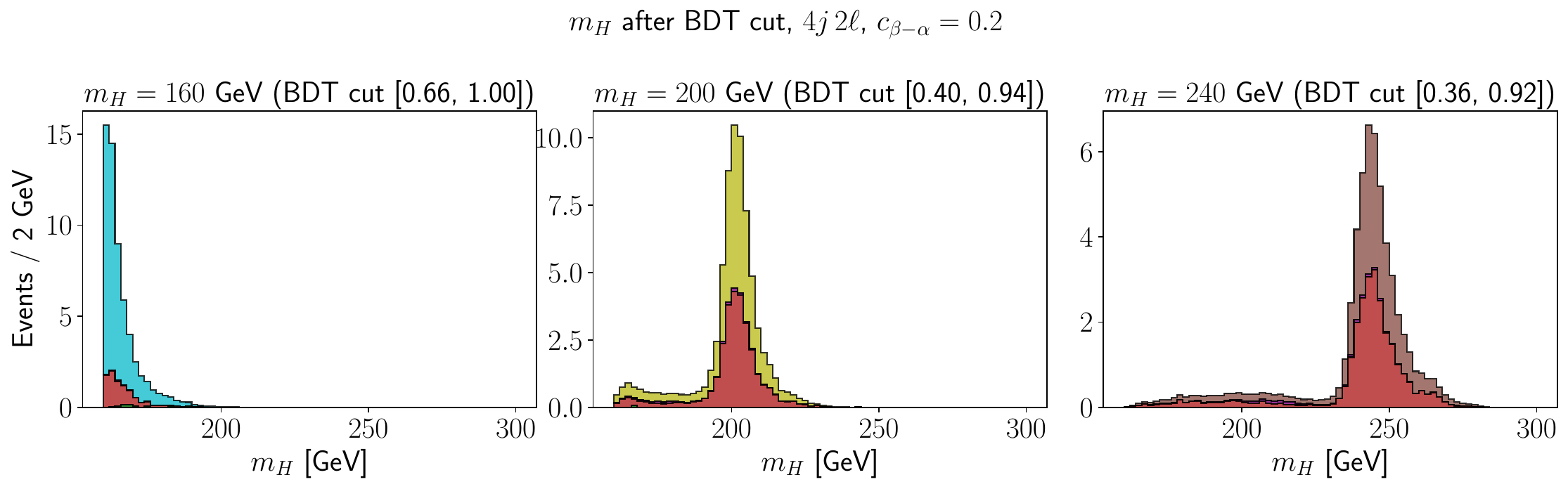}\\
	\includegraphics[width=\textwidth]{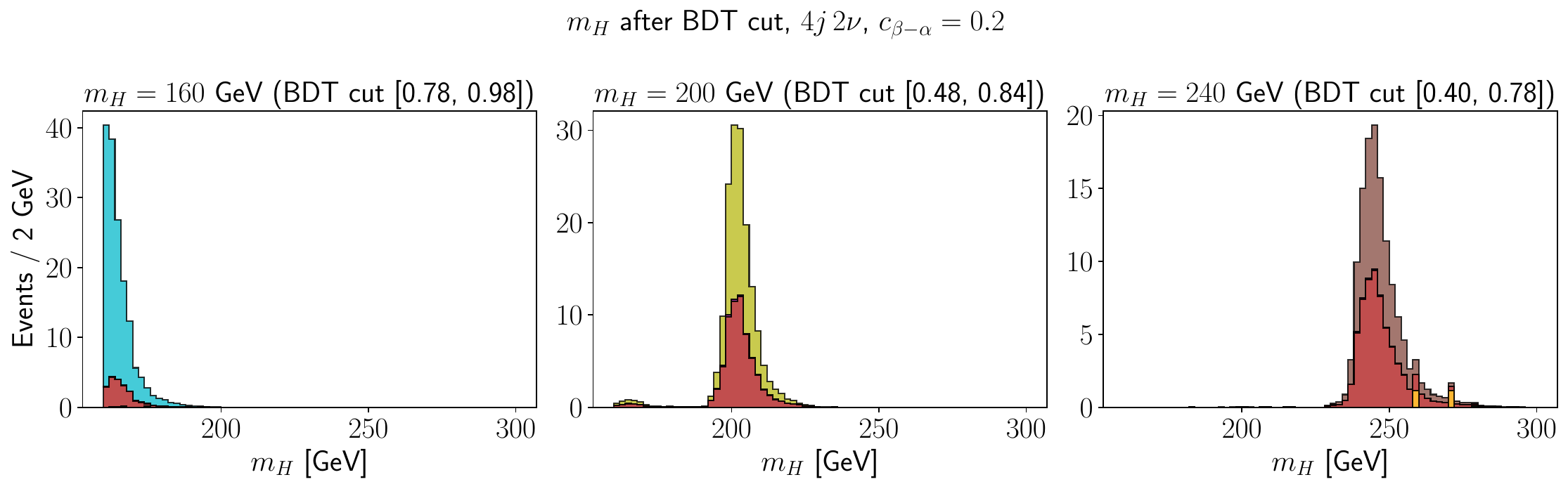}
	\caption{$\mH$ distribution after the optimal 1D BDT cut, for the
		$4j2\ell$ (top) and $4j2\nu$ (bottom) channels.  Each panel
		corresponds to one representative mass point.}
	\label{fig:bdt_mh}
\end{figure}

Figure~\ref{fig:bdt_mh} shows the $\mH$ distribution after the
optimal 1D BDT cut, for the three representative masses.  The signal
peak is visible at the true mass, and the background is strongly
suppressed by the BDT cut.

\begin{figure}[t]
	\centering
	\includegraphics[width=\textwidth]{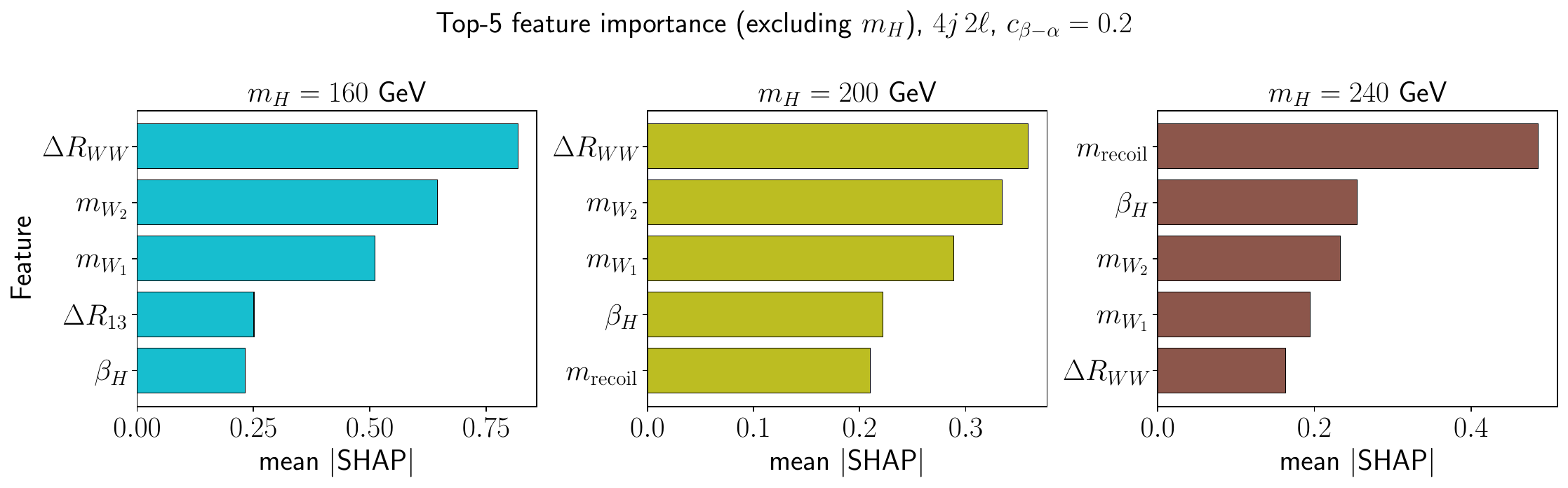}\\
	\includegraphics[width=\textwidth]{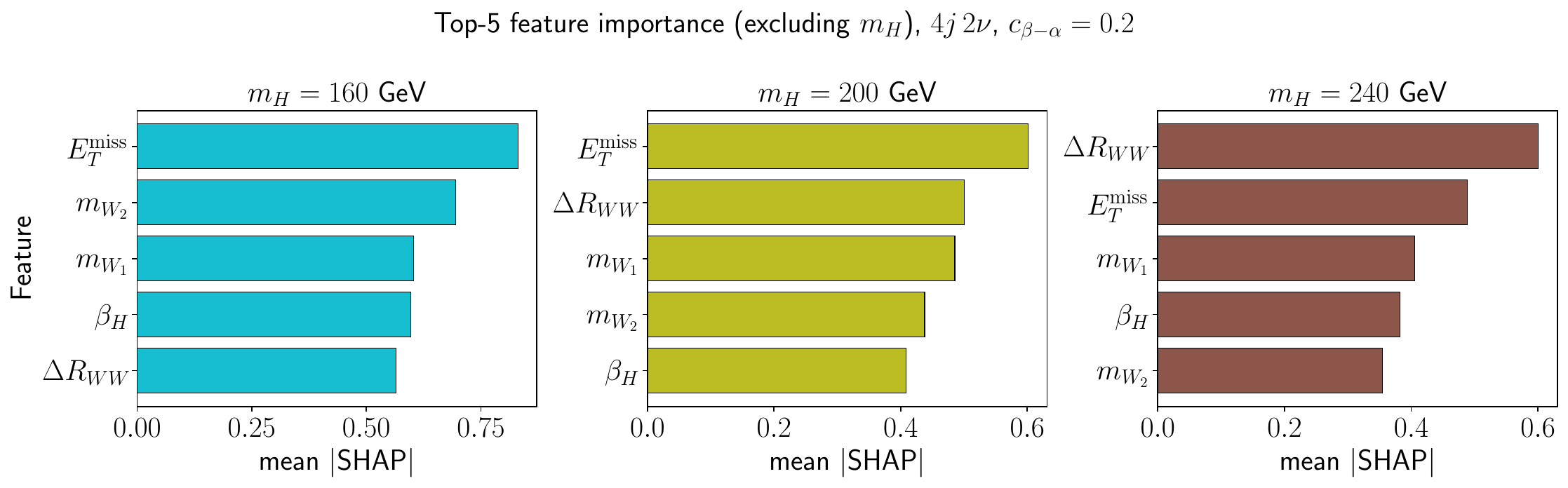}
	\caption{BDT top-five features ranked by mean $|\mathrm{SHAP}|$, for the
		$4j2\ell$ (top) and $4j2\nu$ (bottom) channels.  Each panel
		corresponds to one mass point.}
	\label{fig:shap}
\end{figure}

Figure~\ref{fig:shap} shows the top-five features ranked by their mean
absolute SHAP value~\cite{shap}, separately for each mass point.  The
feature ranking is stable across the mass range: $\Delta R_{WW}$ and
$E_T^{\rm miss}$ dominate at low mass, while the $W$-boson masses and
$m_{\rm recoil}$ become more important at high mass.  The ranking is
consistent between the two channels, which confirms that the
classifier is learning the same physics in both cases.

\subsection{Deep neural network}
\label{sec:dnn}

As an independent cross-check of the BDT we train a fully connected
deep neural network (DNN) on the same feature set and the same
training/testing split, using the $4j2\ell$ channel as an example.
The DNN is a standard fully connected network with three hidden layers
of 128, 64, and 32 units.  As with the BDT, no hyperparameter
optimisation was performed; the architecture is a standard choice and
is not the subject of this paper.

Table~\ref{tab:bdt_dnn} compares the two classifiers at the five mass
points.  The DNN reaches a marginally larger 1D-window significance,
but the difference is below $0.5\sigma$ at every mass point, and does
not change the conclusion of the analysis: both classifiers, with
completely different decision boundaries, give consistent results and
both lie well above the cut-based baseline.  The DNN therefore adds no
sizable sensitivity to the analysis.  It does, however, come at a
substantial computational cost: about twenty times the training time
of the BDT per mass point, as discussed in Sec.~\ref{sec:timing}.
This is the reason the BDT is used as the default classifier
throughout the analysis, with the DNN retained as a cross-check.

\begin{figure}[t]
	\centering
	\includegraphics[width=\textwidth]{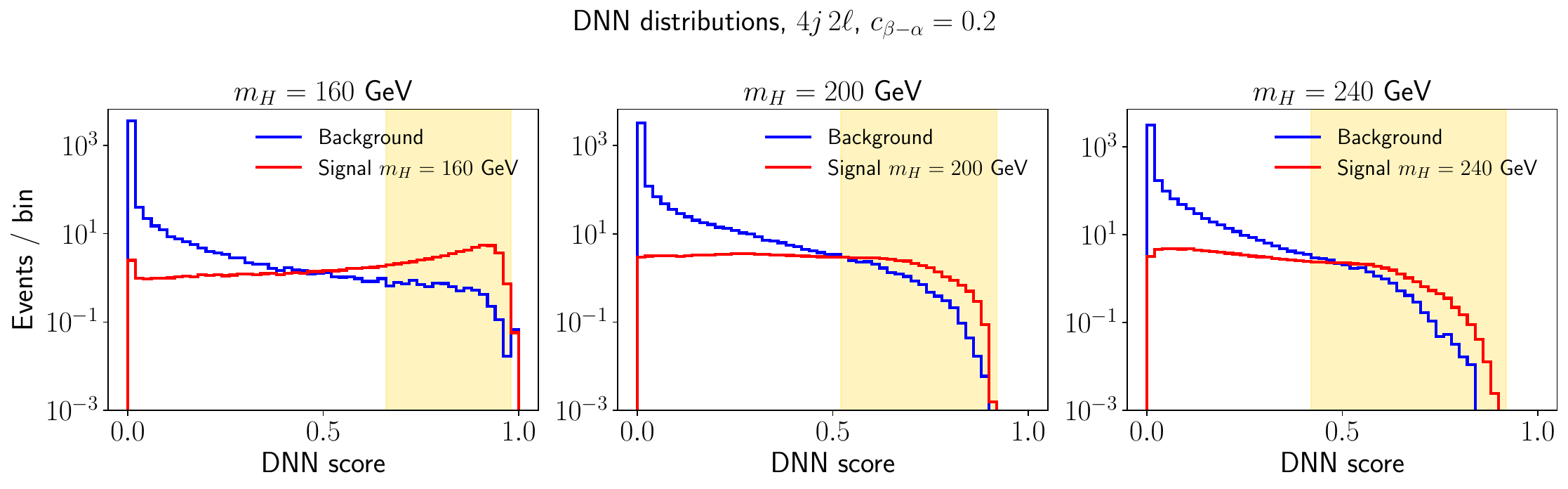}
	\caption{DNN score distributions for background (blue) and signal
		(red) at $\mH=160,200,240~\mathrm{GeV}$ in the $4j2\ell$ channel.
		The gold band indicates the optimal DNN window.}
	\label{fig:dnndist}
\end{figure}

\begin{figure}[t]
	\centering
	\includegraphics[width=\textwidth]{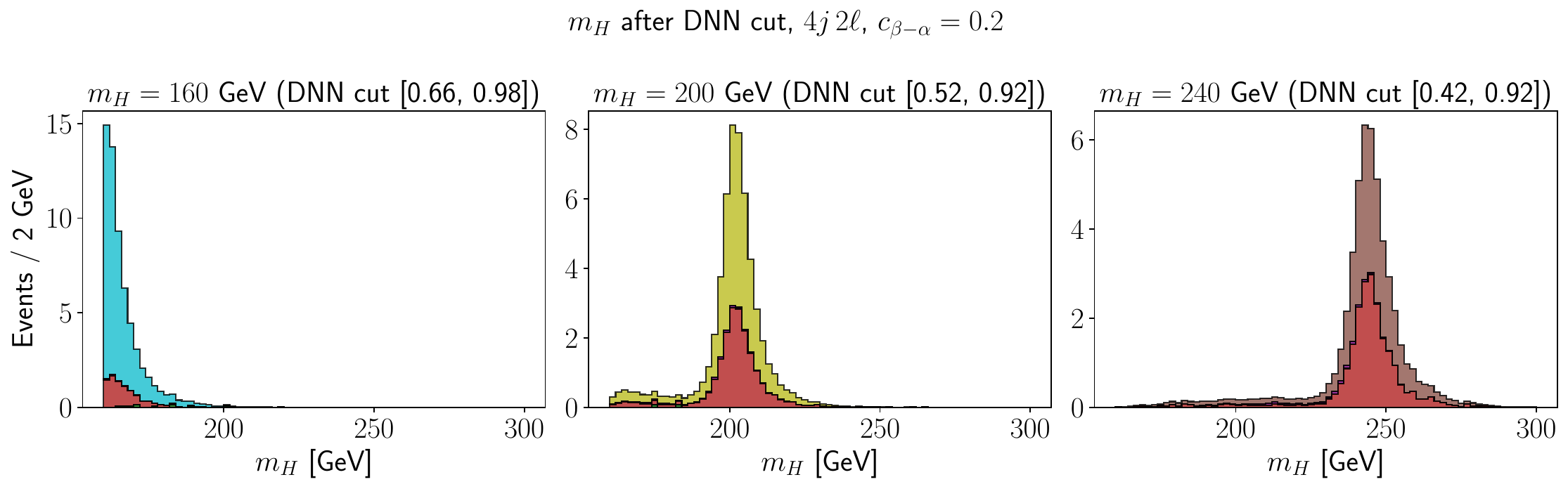}
	\caption{$\mH$ distribution after the optimal DNN cut, in the
		$4j2\ell$ channel.  Each panel corresponds to one representative
		mass point.}
	\label{fig:dnnmh}
\end{figure}

\begin{figure}[t]
	\centering
	\includegraphics[width=\textwidth]{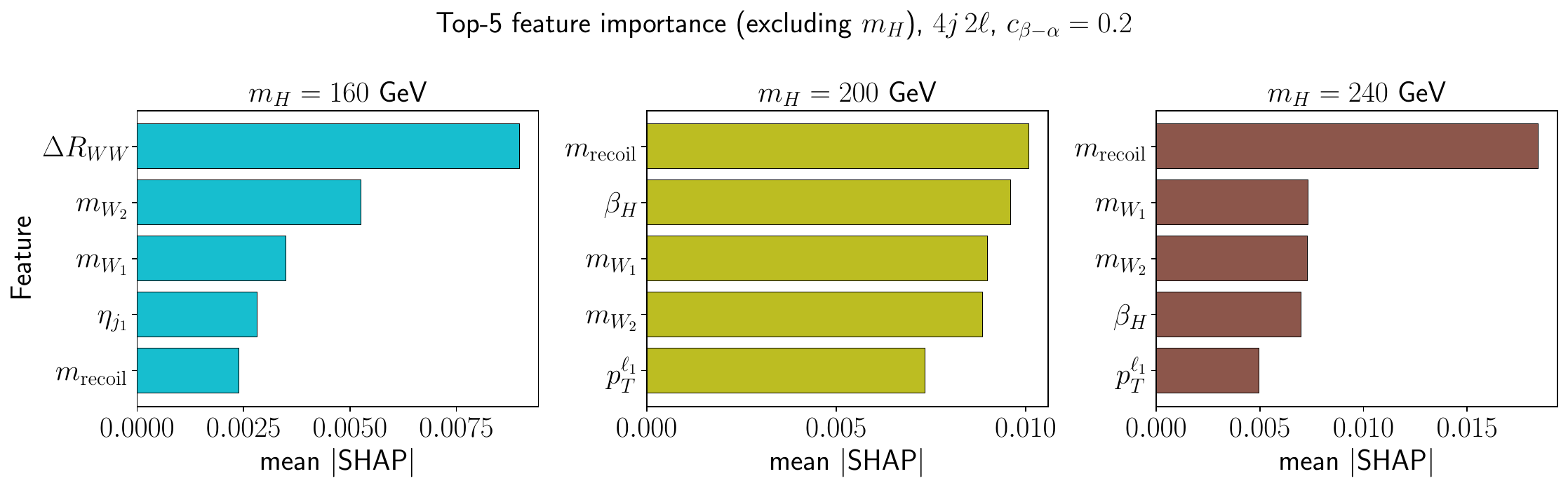}
	\caption{DNN top-five features ranked by mean $|\mathrm{SHAP}|$, in the $4j2\ell$ channel.  Each panel corresponds to one
		mass point.}
	\label{fig:dnnshap}
\end{figure}

Figure~\ref{fig:dnndist} shows the DNN score distributions for
background and signal at the three representative masses, with the
optimal window shaded in gold; Fig.~\ref{fig:dnnmh} shows the $\mH$
distribution after the DNN cut.  Feature importance is computed with
\textsc{shap}~\cite{shap} using the \texttt{DeepExplainer}, with a
background sample of $200$ training events and $500$ test events per
mass point.  Figure~\ref{fig:dnnshap} shows the top-five features.
The ranking is consistent with the BDT (Fig.~\ref{fig:shap}), which
confirms that both classifiers exploit the same physics.
\begin{table}[t]
	\centering
	\caption{
		Comparison of the BDT and DNN classifiers at the five mass
		points, in the $4j2\ell$ channel.  The weighted AUC and the
		1D-window significance $Z_{\rm 1D}$ are shown for each
		classifier.  The DNN was not trained in the $4j2\nu$ channel
		(see text); the comparison with the cut-based and 2D-template
		analyses is given in Tables~\ref{tab:summary-ll}
		and~\ref{tab:summary-nn}.
	}
	\label{tab:bdt_dnn}
	\begin{tabular}{rcccc}
		\toprule
		& \multicolumn{2}{c}{AUC} & \multicolumn{2}{c}{$Z_{\rm 1D}$} \\
		\cmidrule(lr){2-3}\cmidrule(lr){4-5}
		$\mH$ [GeV] & BDT & DNN & BDT & DNN \\
		\midrule
		160 & 0.988 & 0.991 & 10.08 & 10.63 \\
		180 & 0.976 & 0.979 &  6.92 &  7.22 \\
		200 & 0.962 & 0.969 &  5.10 &  5.50 \\
		220 & 0.955 & 0.962 &  4.66 &  5.01 \\
		240 & 0.951 & 0.960 &  4.12 &  4.58 \\
		\bottomrule
	\end{tabular}
\end{table}

	% =====================================================================
	\section{Statistical methodology}
	\label{sec:stats}
	
	\subsection{Asimov significance for a counting experiment}
	\label{sec:asimov}
	
For a counting experiment with $N_S$ signal and $N_B$ background
events, and a Gaussian background uncertainty $\sigma_B = \epsilon_B N_B$
(with $\epsilon_B$ the fractional uncertainty), the profiled Asimov
significance is
\begin{equation}
	Z^2 = 2\left[
	(N_S+N_B)\ln\frac{(N_S+N_B)(N_B+\sigma_B^2)}
	{N_B^2+(N_S+N_B)\sigma_B^2}
	- \frac{N_B^2}{\sigma_B^2}\ln\!\left(1+
	\frac{\sigma_B^2 N_S}{N_B(N_B+\sigma_B^2)}\right)
	\right].
	\label{eq:asimov}
\end{equation}
This is the form used in the cut-based scan, where $N_S$ and $N_B$ are
the yields inside a mass window $[\mH^{\rm lo},\mH^{\rm hi}]$, and in
the 1D BDT scan, where they are the yields inside a BDT window
$[b_1,b_2]$.  In both cases all contiguous windows are scanned and the
one maximising $Z$ is retained.  The fractional uncertainty is taken as
$\epsilon_B=0.10$ for the cut-based and 1D BDT scans, and
$\epsilon_B=\sqrt{0.10^2+\sigma_{\rm JES}^2}$ for the 2D template (see
Sec.~\ref{sec:jes}).
	
	\subsection{Profile likelihood}
	\label{sec:pyhf}
	
	For the 2D $(\mathrm{BDT},\mH)$ template the significance is computed
	with the profile-likelihood-ratio statistic
	\begin{equation}
		q_0 = -2\ln\frac{L(\mu=0,\hat{\hat\theta}_0)}
		{L(\hat\mu,\hat\theta)},
	\end{equation}
	evaluated on the Asimov dataset, using the \textsc{pyhf}
	implementation~\cite{pyhf}.  The signal strength $\mu$ is the
	parameter of interest; the background normalisation and the JES
	uncertainty are treated as Gaussian nuisance parameters.  The
	significance is $\Ztwod = \sqrt{q_0}$.  This is the quantity reported
	in the $Z_{\rm 2D}$ column of Tables~\ref{tab:summary-ll}
	and~\ref{tab:summary-nn}.
	
	\subsection{CL$_s$ exclusion contours}
	\label{sec:cls}
	
	For the exclusion contours we use the modified frequentist
	$\CLs$ construction~\cite{Cowan:2010js} with the $\tilde q_\mu$ test
	statistic, and compute the expected $\CLs$ at each point of the
	$(\cba,\tb)$ grid.  The grid contains $1010$ points for each mass and
	channel, sampled with a stride that covers the relevant range of
	$\cba\in[-0.3,0.3]$ and $\tb\in[1,10]$.  The $95\%$ CL contour is
	obtained by interpolation of the $\CLs$ values on the grid.
	
	\subsection{JES uncertainty}
	\label{sec:jes}
	
	The jet energy scale uncertainty is propagated by shifting the jet
	four-momenta by $\pm 1\sigma$ and recomputing the 2D template.  The
	fractional shift of the total background yield is
	\begin{equation}
		\delta_{\rm JES}^{\pm} = \frac{N_B^{\pm}-N_B}{N_B},
	\end{equation}
	and the symmetrised uncertainty is
	$\sigma_{\rm JES} = \tfrac12|\delta_{\rm JES}^{+}-\delta_{\rm JES}^{-}|$.
	The total background uncertainty entering Eq.~(\ref{eq:asimov}) is
	then $\epsilon_B = \sqrt{0.10^2+\sigma_{\rm JES}^2}$.  The resulting
	values are $\epsilon_B=0.101$ for the $4j2\ell$ channel and
	$\epsilon_B=0.105$ for the $4j2\nu$ channel, with
	$\sigma_{\rm JES}=0.015$ and $0.032$, respectively.
	
	% =====================================================================
	\section{Results}
	\label{sec:results}
\subsection{Per-mass comparison of the three strategies}
\label{sec:z2d}
\begin{figure}[t]
	\centering
	\includegraphics[width=0.48\textwidth]{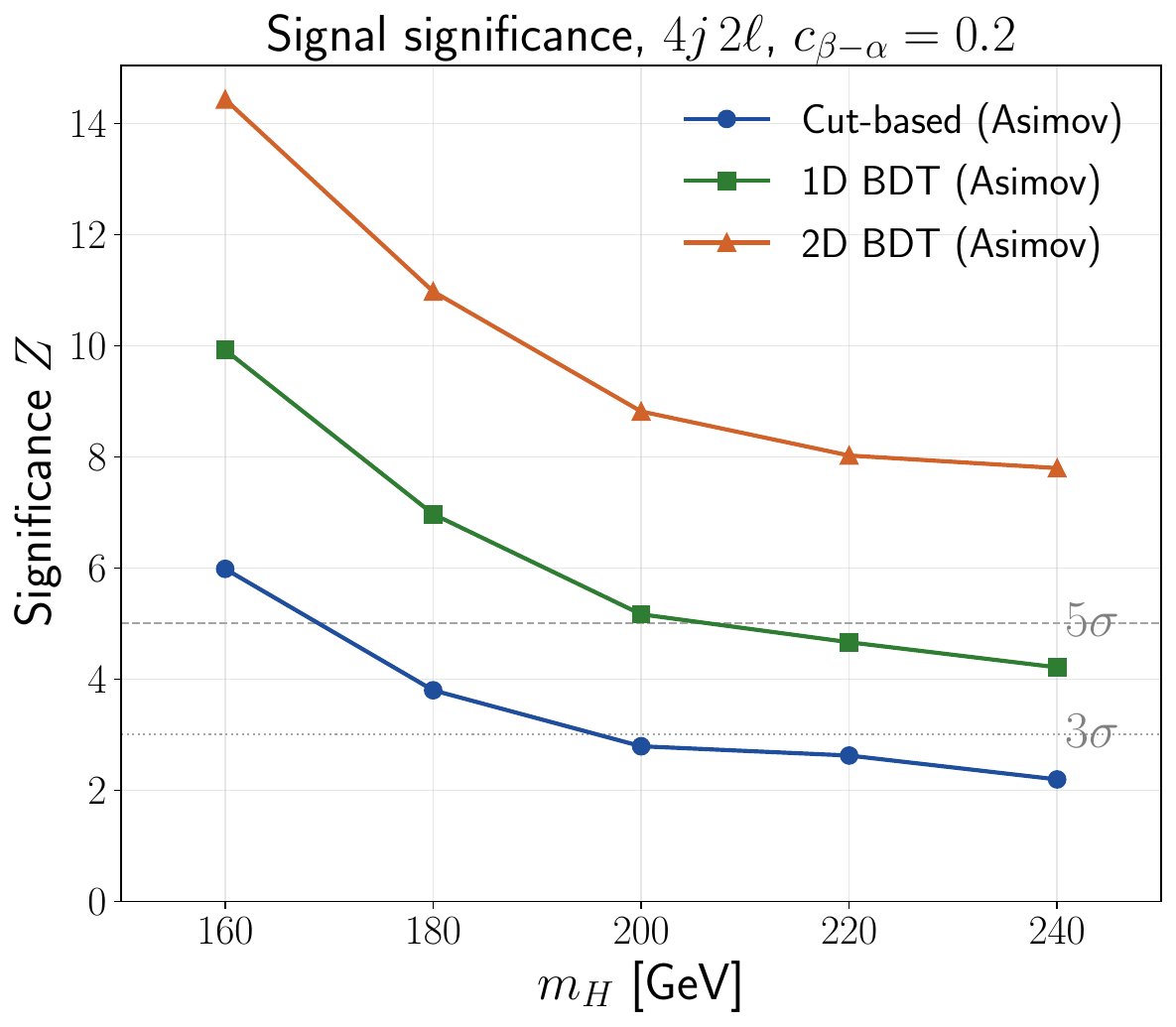}
	\includegraphics[width=0.48\textwidth]{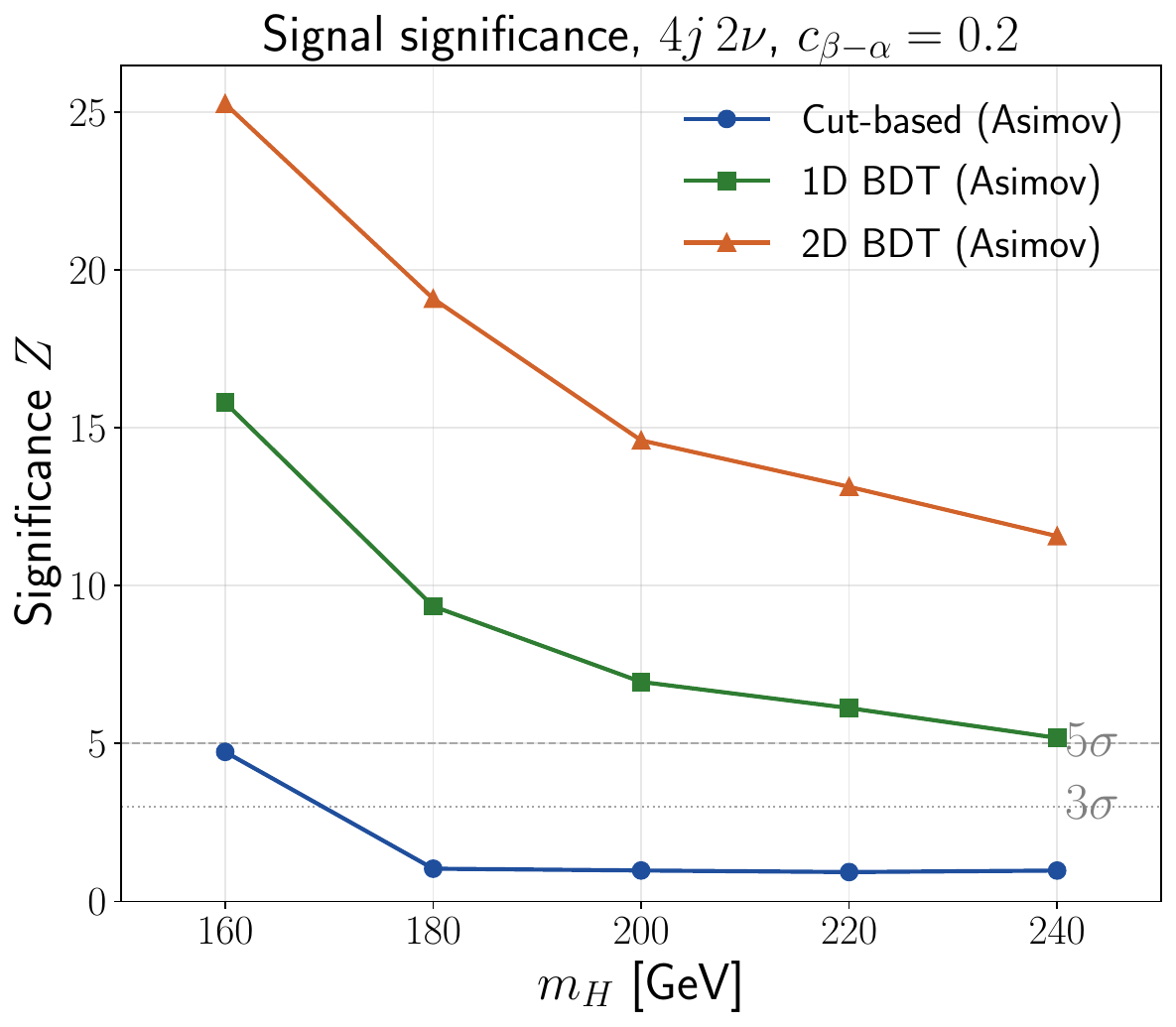}
	\caption{
		Expected signal significance $Z$ as a function of $\mH$ for the
		cut-based analysis (blue circles), the 1D BDT-window scan (green
		squares), and the 2D $(\mathrm{BDT},\mH)$ template (orange
		triangles), for the $4j2\ell$ (left) and $4j2\nu$ (right)
		channels at $(\cba,\tb)=(0.2,5)$.  The dashed and dotted grey
		lines mark the $5\sigma$ and $3\sigma$ thresholds.  The
		corresponding numerical values are given in
		Tables~\ref{tab:summary-ll} and~\ref{tab:summary-nn}.
	}
	\label{fig:significance}
\end{figure}
Tables~\ref{tab:summary-ll} and~\ref{tab:summary-nn} collect the
per-mass results for the three strategies — the cut-based analysis
($Z_{\rm cut}$), the 1D BDT-window scan ($Z_{\rm 1D}$), and the 2D
$(\mathrm{BDT},\mH)$ template ($Z_{\rm 2D}$) — together with the
ratios that quantify the gain of the ML strategies over the cut-based
baseline.  The reference point is $(\cba,\tb)=(0.2,5)$ for all three
columns; the 2D column is the quantity defined in
Sec.~\ref{sec:pyhf}.
Figure~\ref{fig:significance} shows the same results graphically,
with the $3\sigma$ and $5\sigma$ discovery thresholds marked.

The 1D BDT improves the expected significance by a factor of
$1.68$--$1.88$ in the $4j2\ell$ channel and by a factor of
$3.36$--$8.91$ in the $4j2\nu$ channel, with the larger gains at high
mass where the cut-based mass window loses efficiency.  The
improvement is driven by the same physics in both channels: the BDT
exploits the correlation between the BDT score and the reconstructed
$\mH$, which a sequential cut cannot access.  The effect is
particularly pronounced in the $4j2\nu$ channel, where the
preselection is deliberately loose (a lepton veto plus a missing
momentum requirement) and the cut-based analysis therefore loses a
large fraction of the signal at high mass.

\begin{figure}[htbp]
	\centering
	\includegraphics[width=\textwidth]{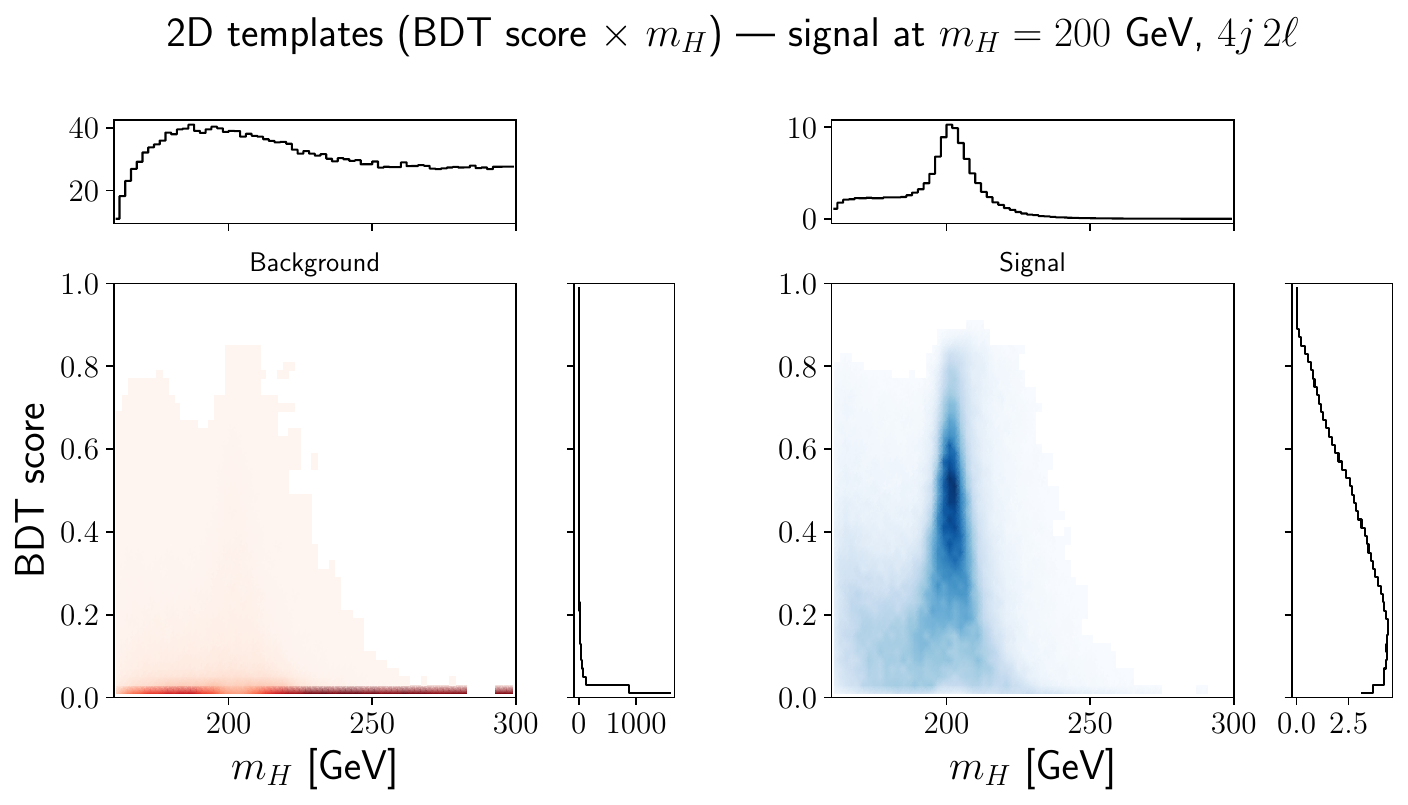}\\
	\includegraphics[width=\textwidth]{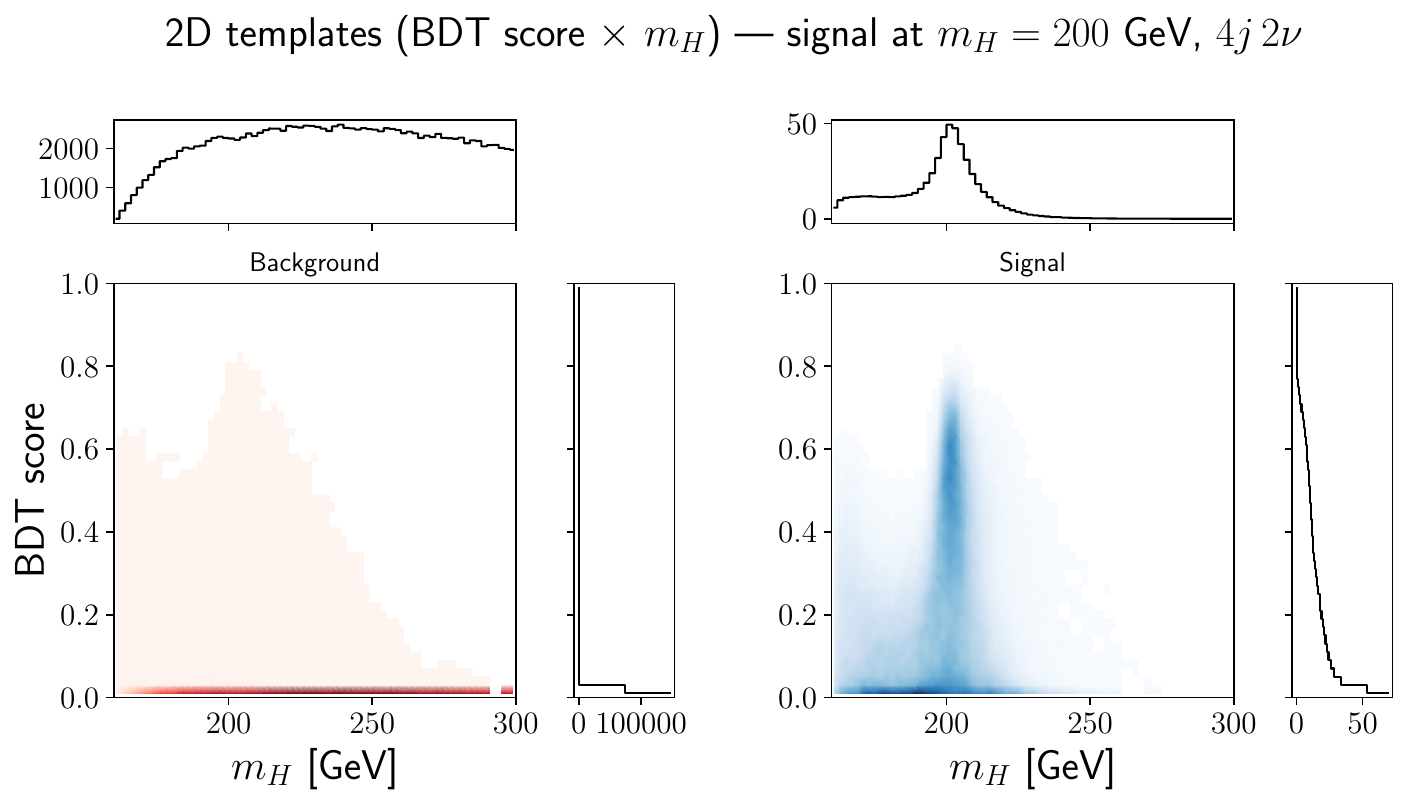}
	\caption{2D $(\mathrm{BDT},\mH)$ templates for background (left) and
		signal (right) at $\mH=200~\mathrm{GeV}$, for the $4j2\ell$ (top)
		and $4j2\nu$ (bottom) channels.  Each panel shows the 2D
		distribution together with the two 1D marginals.}
	\label{fig:2dtemplates}
\end{figure}

The 2D template goes one step further.  Figure~\ref{fig:2dtemplates}
shows the 2D $(\mathrm{BDT},\mH)$ templates for the background and for
a representative signal mass ($\mH=200~\mathrm{GeV}$) in the
$4j2\ell$ channel.  The background template is concentrated at low BDT
score and covers the full mass range, whereas the signal template is
concentrated at high BDT score and at the reconstructed mass peak.
The 2D likelihood exploits both features simultaneously, which is why
it outperforms the 1D BDT scan: the 1D scan collapses the
$(\mathrm{BDT},\mH)$ distribution to a single BDT window, whereas the
2D template retains the full shape information.  The resulting gain
over the cut-based baseline is a factor of $2.45$--$3.57$ in the
$4j2\ell$ channel and $5.34$--$18.34$ in the $4j2\nu$ channel.  As
for the 1D BDT, the improvement grows with $\mH$ in both channels,
which is consistent with the degradation of the cut-based selection
at high mass.

\begin{table}[t]
	\centering
	\caption{Per-mass summary for the $4j2\ell$ channel at the reference
		point $(\cba,\tb)=(0.2,5)$.  $Z_{\rm cut}$, $Z_{\rm 1D}$, and
		$Z_{\rm 2D}$ are the cut-based, 1D BDT, and 2D template
		significances, respectively; the two rightmost columns quantify
		the gain of the ML strategies over the cut-based baseline.}
	\label{tab:summary-ll}
	\begin{tabular}{rrrrrr}
		\toprule
		$\mH$ [GeV] & $Z_{\rm cut}$ & $Z_{\rm 1D}$ & $Z_{\rm 2D}$ & $Z_{\rm 1D}/Z_{\rm cut}$ & $Z_{\rm 2D}/Z_{\rm cut}$ \\
		\midrule
		160 & 5.98 & 10.08 & 14.65 & 1.68 & 2.45 \\
		180 & 3.80 &  6.92 & 10.84 & 1.82 & 2.85 \\
		200 & 2.79 &  5.10 &  8.69 & 1.83 & 3.11 \\
		220 & 2.62 &  4.66 &  8.08 & 1.77 & 3.08 \\
		240 & 2.20 &  4.12 &  7.83 & 1.88 & 3.57 \\
		\bottomrule
	\end{tabular}
\end{table}

\begin{table}[t]
	\centering
	\caption{Same as Table~\ref{tab:summary-ll}, for the $4j2\nu$
		channel.}
	\label{tab:summary-nn}
	\begin{tabular}{rrrrrr}
		\toprule
		$\mH$ [GeV] & $Z_{\rm cut}$ & $Z_{\rm 1D}$ & $Z_{\rm 2D}$ & $Z_{\rm 1D}/Z_{\rm cut}$ & $Z_{\rm 2D}/Z_{\rm cut}$ \\
		\midrule
		160 & 4.73 & 15.91 & 25.30 & 3.36 &  5.34 \\
		180 & 1.03 &  9.22 & 18.98 & 8.91 & 18.34 \\
		200 & 0.98 &  6.83 & 14.39 & 6.99 & 14.72 \\
		220 & 0.93 &  6.03 & 13.02 & 6.49 & 14.03 \\
		240 & 0.98 &  5.28 & 11.53 & 5.41 & 11.82 \\
		\bottomrule
	\end{tabular}
\end{table}

	\subsection{Exclusion contours in the $(\cba,\tb)$ plane}
	\label{sec:contours}
	
\begin{figure}[htbp]
		\centering
		\includegraphics[width=0.48\textwidth]{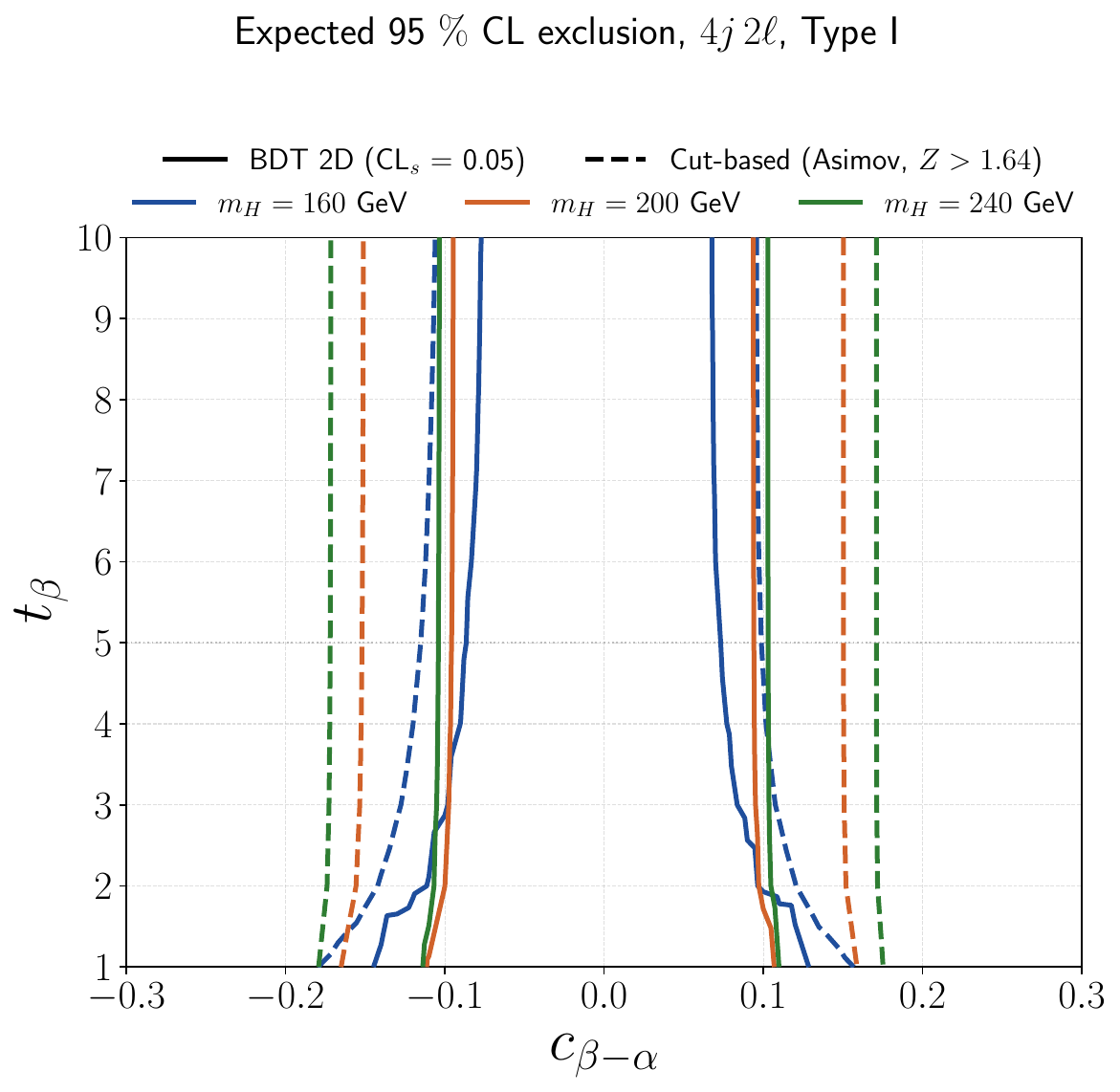}
		\includegraphics[width=0.48\textwidth]{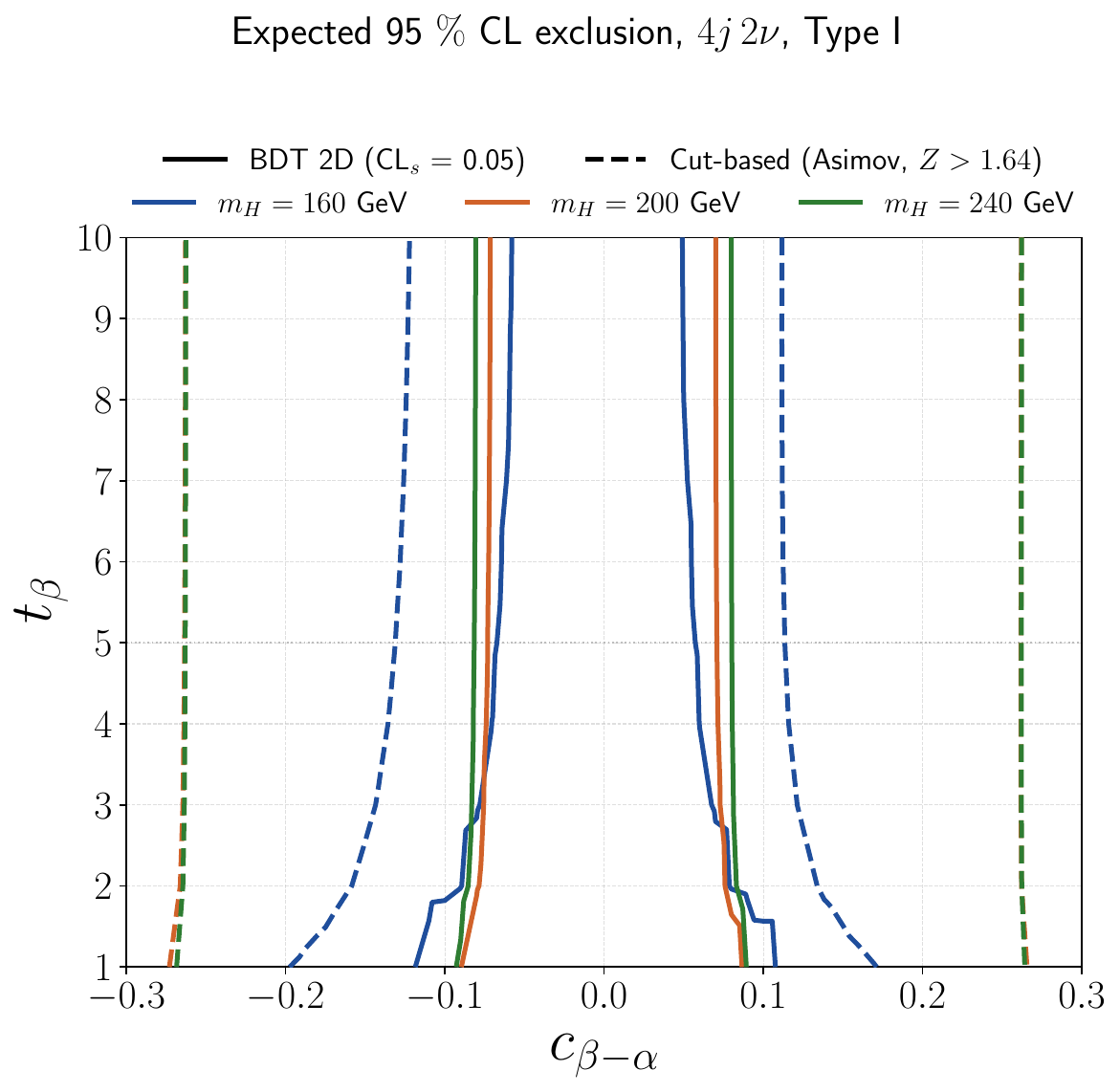}
		\caption{Expected $95\%$ CL exclusion contours in the
			$(\cba,\tb)$ plane for the $4j2\ell$ (left) and $4j2\nu$ (right)
			channels.  Solid lines: 2D template ($\CLs=0.05$); dashed lines:
			cut-based analysis (Asimov, $Z>1.64$).  Colours distinguish the
			mass points.}
		\label{fig:cls}
	\end{figure}

	Figure~\ref{fig:cls} shows the expected $95\%$ CL exclusion contours
	in the $(\cba,\tb)$ plane, for the three representative masses
	$\mH=160,200,240~\mathrm{GeV}$ and for both channels.  Solid lines
	correspond to the 2D template, dashed lines to the cut-based analysis.
	The 2D template systematically excludes a larger region of parameter
	space, and the gap between the two contours is larger in the $4j2\nu$
	channel, consistent with the significance gain in
	Tables~\ref{tab:summary-ll} and~\ref{tab:summary-nn}.
	
	\subsection{$\sbr$ upper limits}
	\label{sec:sigbr}
	
\begin{figure}[htbp]
		\centering
		\includegraphics[width=0.48\textwidth]{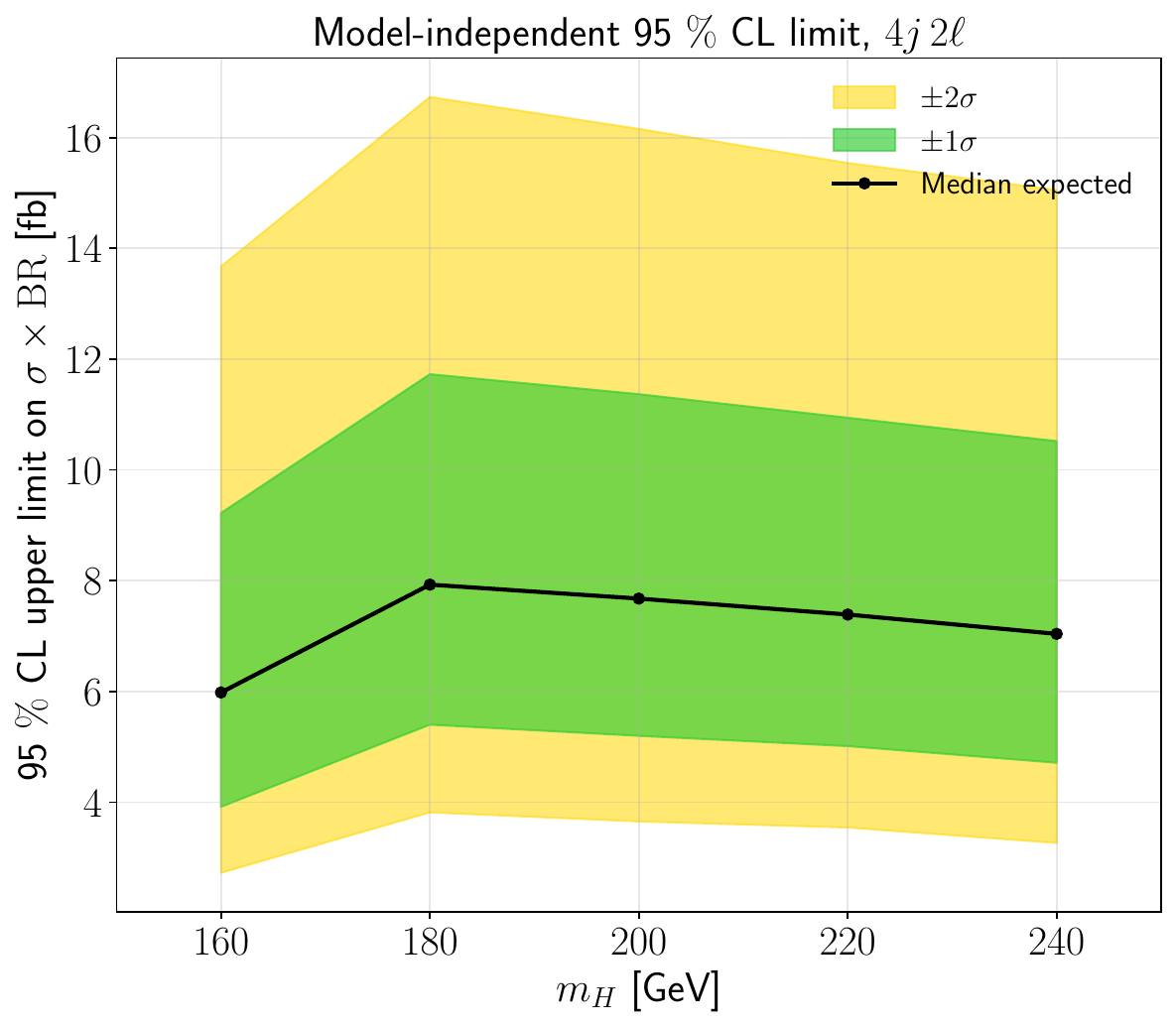}
		\includegraphics[width=0.48\textwidth]{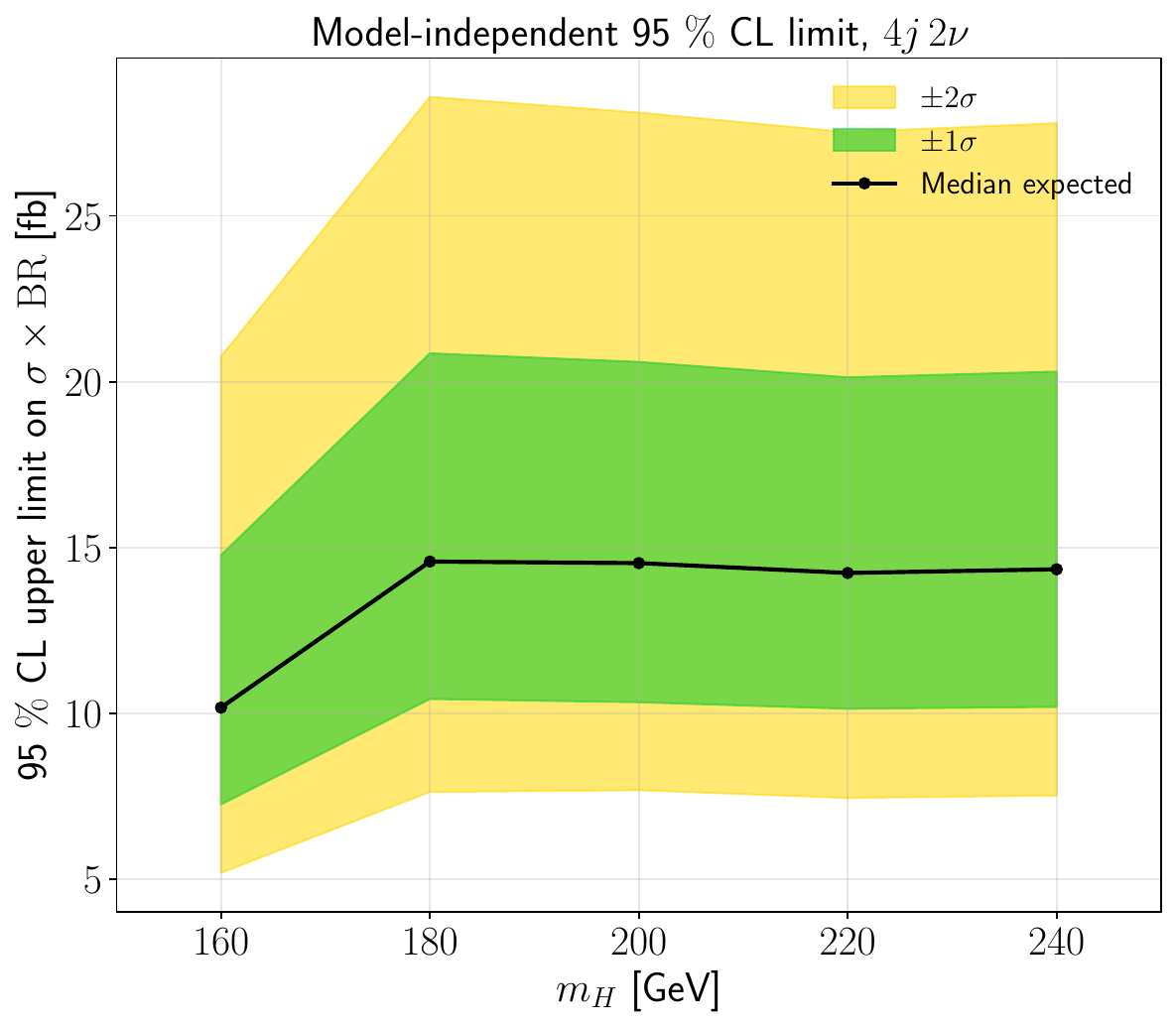}
		\caption{$95\%$ CL upper limits on $\sbr$ as a function of $\mH$ for
			the $4j2\ell$ (left) and $4j2\nu$ (right) channels.  The black
			line is the median expected limit; the green and gold bands
			indicate the $\pm1\sigma$ and $\pm2\sigma$ uncertainties.}
		\label{fig:sigbr}
	\end{figure}

	Figure~\ref{fig:sigbr} shows the $95\%$ CL upper limits on
	$\sbr$ as a function of $\mH$, obtained from a $\mu$ scan with the
	$\CLs$ construction.  The median expected limits and
	the $\pm1\sigma$ and $\pm2\sigma$ bands are shown.  In the
	$4j2\ell$ channel the median limit is $5.9$--$8.1~\mathrm{fb}$ over
	the mass range $160$--$240~\mathrm{GeV}$; in the $4j2\nu$ channel it
	is $10.2$--$14.8~\mathrm{fb}$.  The limits are weakest around
	$\mH\sim 180$--$200~\mathrm{GeV}$, where the $WW$ and $ZZ$ backgrounds
	peak.
	
\subsection{Timing}
\label{sec:timing}

The DNN cross-check in Sec.~\ref{sec:dnn} showed no sizable gain over
the BDT; the computational cost is the other half of the
BDT-versus-DNN comparison. 
Table~\ref{tab:timing} compares the training time of the two
classifiers in the $4j2\ell$ channel, per mass point.  The DNN takes
roughly twenty times longer to train than the BDT: the BDT is
essentially a single pass over the sorted features (with histogram
binning, \texttt{tree\_method="hist"}), while the DNN performs
$50$ epochs of batched gradient descent with early stopping, each
epoch requiring a forward and backward pass over the full training
set.  The difference is not a problem in practice — even the DNN
trains in under four minutes per mass point — but it is worth stating
explicitly, because it is the main computational cost of choosing a
DNN over a BDT for this analysis.

The $\CLs$ scan is the other significant cost: each
mass point requires $1010$ evaluations of the $\tilde q_\mu$ test
statistic, one per grid point, for each of the three mass points used
in the exclusion contours. The event generation chain
(\textsc{WHIZARD} + \textsc{Pythia} + \textsc{Delphes}) is the
dominant cost overall; the ML training and the statistical scan are
subdominant.

\begin{table}[t]
	\centering
	\caption{Training time per mass point for the BDT and the DNN, in
		the $4j2\ell$ channel.  The values are the mean over the five
		mass points ($160$--$240~\mathrm{GeV}$) quoted in the run log;
		the DNN time includes the early-stopping validation pass.  This
		is training time only — scoring the full event sample takes a
		few seconds for both classifiers and is not included.}
	\label{tab:timing}
	\begin{tabular}{lcc}
		\toprule
		Classifier & Training time per mass [s] & Ratio to BDT \\
		\midrule
		XGBoost (BDT) & $11.5 \pm 0.4$ & $1$ \\
		DNN           & $233 \pm 7$    & $\approx 20$ \\
		\bottomrule
	\end{tabular}
\end{table}
	
	% =====================================================================
	\section{Discussion and caveats}
	\label{sec:discussion}
	
The improvement of the 2D template over the cut-based analysis has
three sources.  First, the BDT itself exploits correlations that are
invisible to sequential cuts.  Second, the 2D template uses the full
shape of the $(\mathrm{BDT},\mH)$ distribution, whereas the 1D scan
collapses it to a single window.  Third, the 2D template uses the full
shape information in the profile-likelihood fit, whereas the 1D scan
retains only the event counts in the mass window.

Several caveats apply.  The analysis is at leading order in electroweak
and QCD corrections, with the signal and background normalised to the
best available cross sections at $\sqrt{s}=500~\mathrm{GeV}$; a full
NLO electroweak calculation would be desirable.  The JES uncertainty is
propagated as a normalisation-only shift; a shape-dependent uncertainty
would weaken the 2D template more than the cut-based analysis, since the
template uses the shape information.

The cut-based baseline used here is a re-implementation of the selection of
Ref.~\cite{Hashemi2026}; it is also based on an optimised mass window,
but it differs from the published lepton-collider result of that paper
in that the latter uses the simple Asimov formula with no systematic
uncertainty, whereas here the profiled Asimov significance of
Eq.~(\ref{eq:asimov}) is used with a $10\%$ background uncertainty.
The two should not be compared directly.  The present analysis is
intended to quantify the ML gain on the same event sample.

The systematic uncertainty budget is deliberately simplified.  The
simulation includes beamstrahlung, jet energy resolution, and ISR/FSR
effects.  A jet energy scale study was also performed, with the JES uncertainty propagated as a normalisation-only shift.  The total background
normalisation uncertainty was obtained by
adding the JES normalisation shift and $10\%$ for the other relevant sources in quadrature.  There is nevertheless open room for a more detailed
treatment: a shape-dependent jet energy scale uncertainty, flavour
tagging, luminosity and beam-energy uncertainties, and theory
uncertainties on the cross sections and scales are not included,
because the study relies on fast simulation, has no access to real
data, and is scoped to isolate the ML gain on a common event sample.
A full systematic programme, ideally on real or fully simulated data,
would be needed to establish the absolute sensitivity and could affect
the relative ranking of the 2D template and the cut-based baseline,
particularly through shape-dependent uncertainties that the template is
more sensitive to.
	% =====================================================================
	\section{Conclusions}
	\label{sec:conclusions}
	
	We have compared a cut-based analysis with two machine-learning
	strategies for the discovery of a heavy CP-even Higgs boson in the
	Type-I 2HDM, in the $4j2\ell$ and $4j2\nu$ final states at a future
	$e^+e^-$ collider with $\sqrt{s}=500~\mathrm{GeV}$.  A 1D BDT
	per mass point improves the expected significance by a factor of
	$1.7$--$1.9$ in the $4j2\ell$ channel and $3.4$--$8.9$ in the
	$4j2\nu$ channel.  A 2D $(\mathrm{BDT},\mH)$ template analysed with a
	profile likelihood improves it by a factor of $2.5$--$3.6$ and
	$5.3$--$18.3$, respectively.  The corresponding $95\%$ CL upper
	limits on $\sbr$ are $5.9$--$8.1~\mathrm{fb}$ ($4j2\ell$) and
	$10.2$--$14.8~\mathrm{fb}$ ($4j2\nu$) over the mass range
	$160$--$240~\mathrm{GeV}$.
	
	The analysis demonstrates that the gain from machine learning in this
	channel is not a small perturbative correction but a factor of
	several, and that it grows with the Higgs mass precisely where the
	cut-based analysis loses efficiency.
	
	% =====================================================================
\section*{Acknowledgements}
We are grateful to the College of Science for providing computing
facilities.
	% =====================================================================
	\appendix
	\section{JES validation and feature distributions}
	\label{app:jes}
	
\begin{figure}[htbp]
		\centering
		\includegraphics[height=0.4\textheight,width=0.9\textwidth]{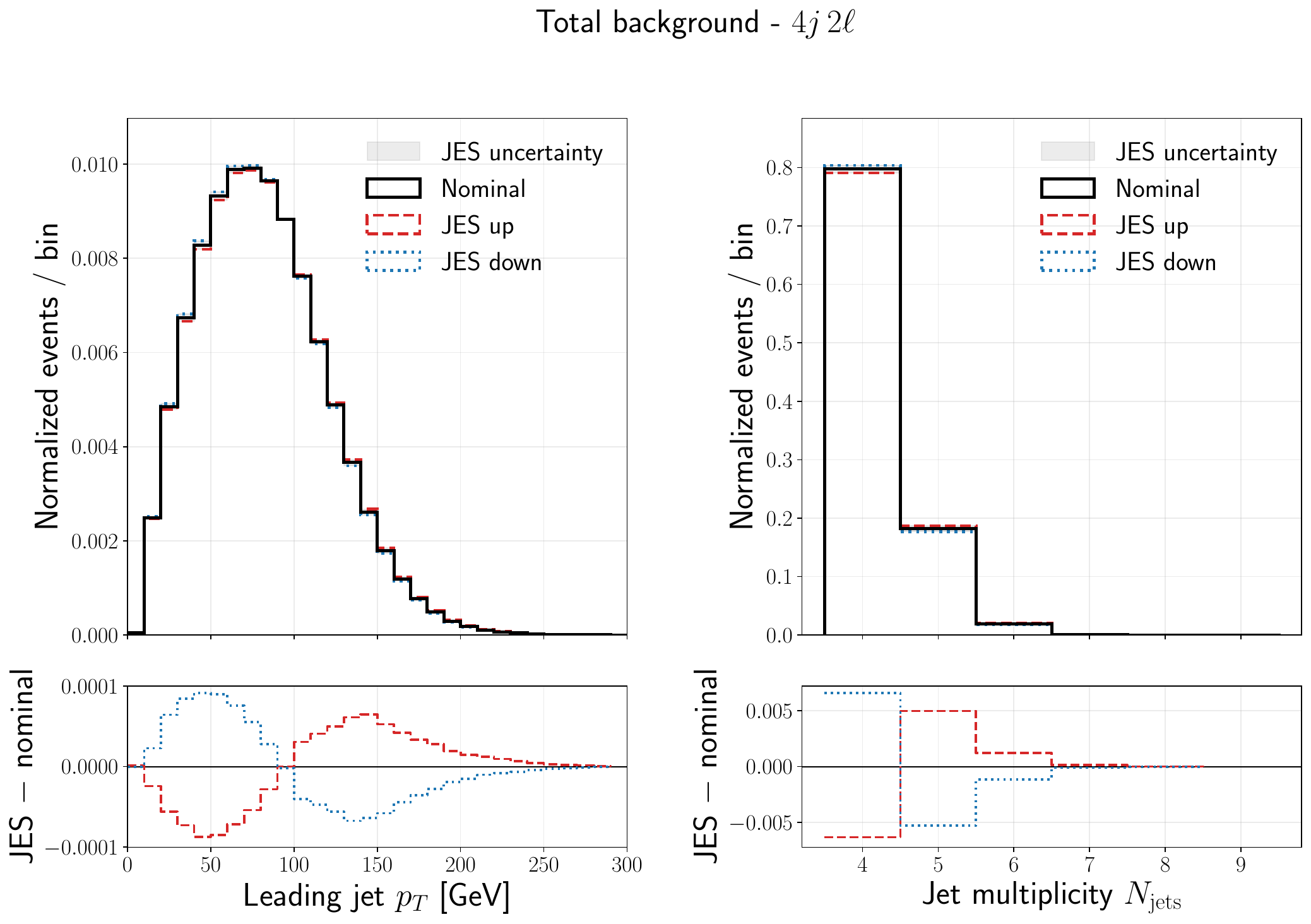}\\
		\includegraphics[height=0.4\textheight,width=0.9\textwidth]{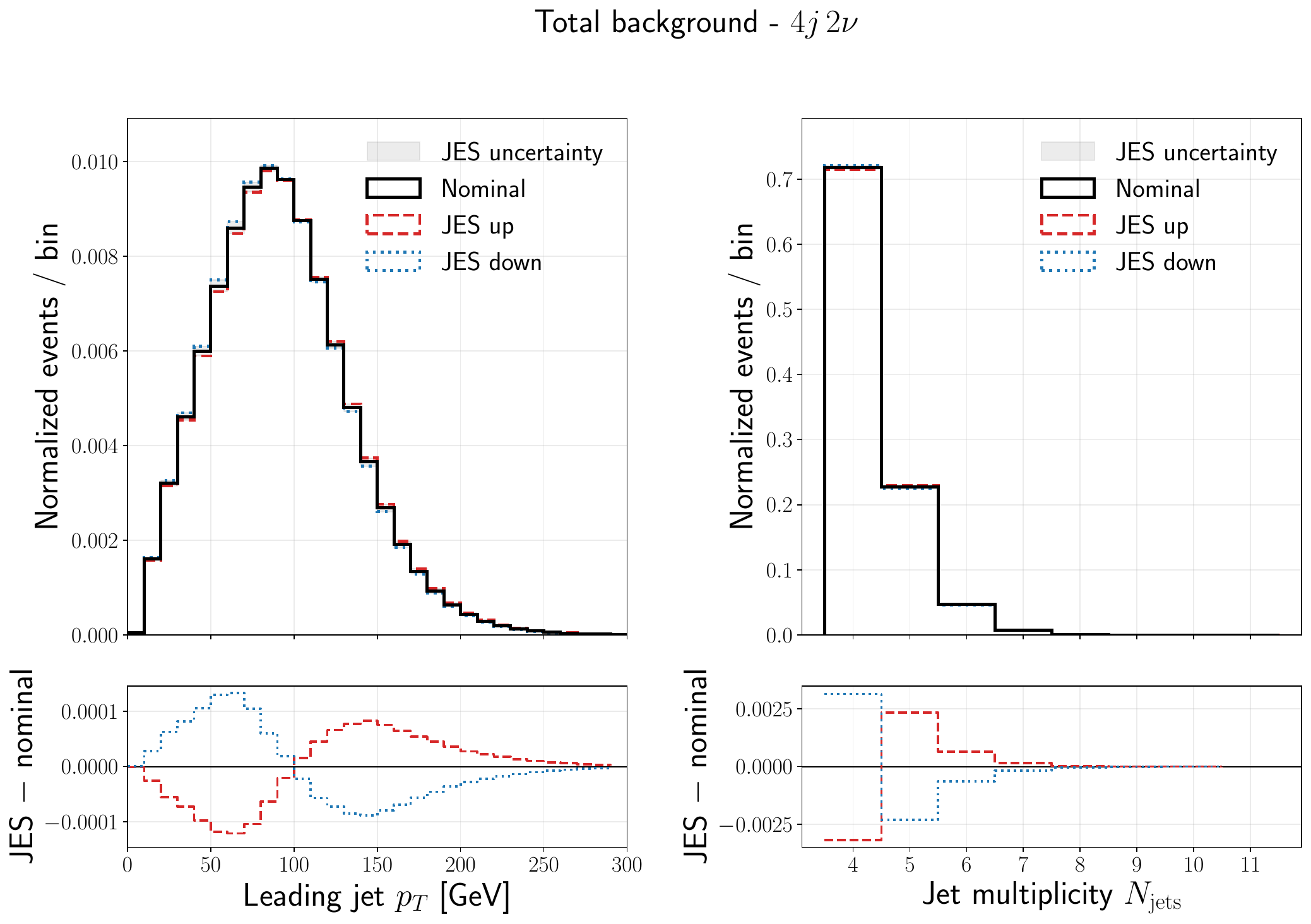}\\
		\caption{JES validation for the $4j2\ell$ channel (top) and $4j2\nu$ (bottom). The leading-jet	$p_T$ is draw on the left column and jet multiplicity on the right, for the nominal,
			JES-up, and JES-down samples.  The lower panels show the
			fractional difference from the nominal.}
		\label{fig:jes}
	\end{figure}

	Figure~\ref{fig:jes} shows the JES validation plots: the leading-jet
	$p_T$ and the jet multiplicity in the nominal, JES-up, and JES-down
	samples, together with the fractional difference from the nominal.
	The JES shift is approximately $\pm1.5\%$ in the $4j2\ell$ channel and
	$\pm3.2\%$ in the $4j2\nu$ channel, consistent with the numbers
	quoted in Sec.~\ref{sec:jes}.
	
\begin{figure}[htbp]
		\centering
		\includegraphics[width=\textwidth]{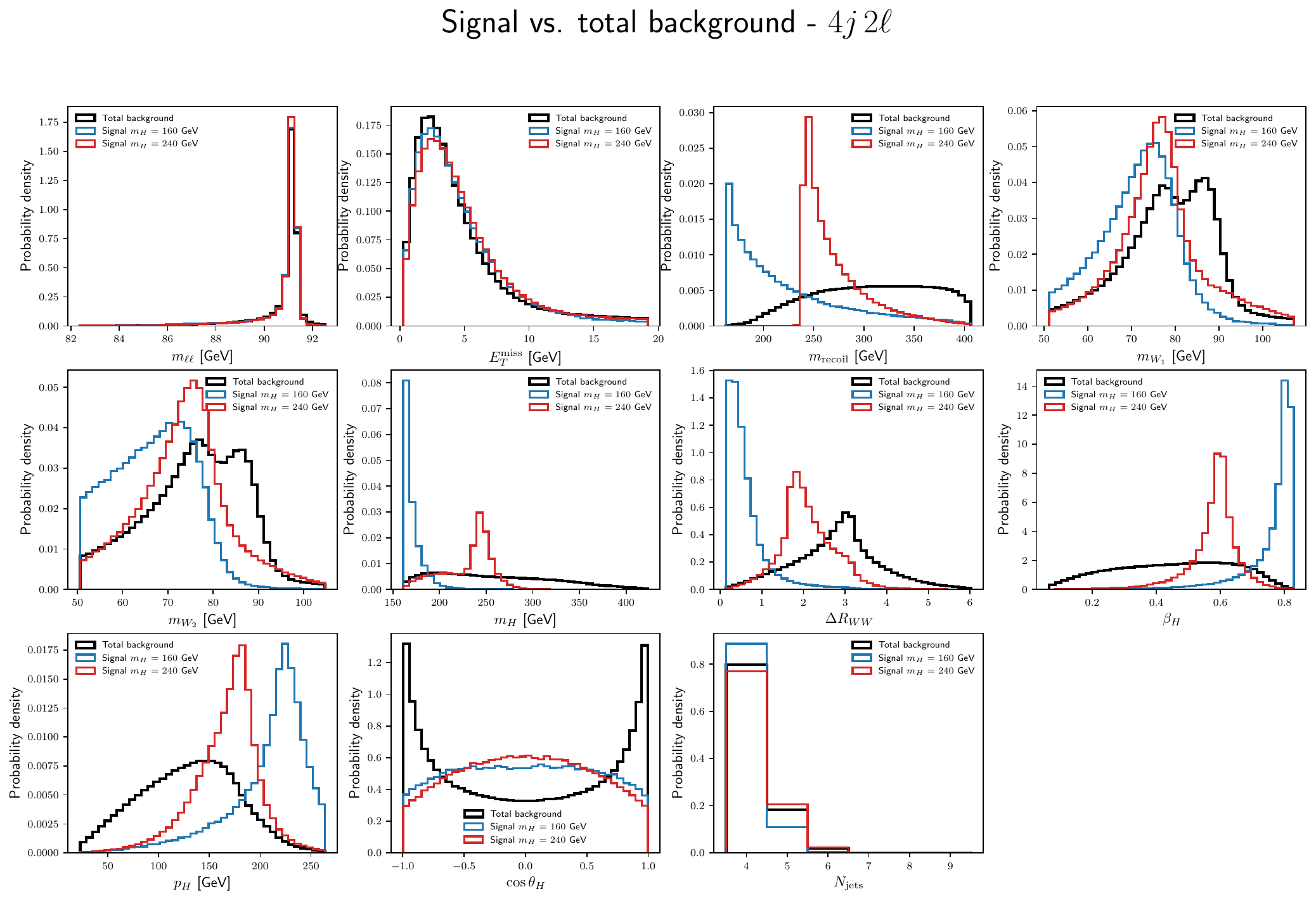}\\[2pt]
		\includegraphics[width=0.8\textwidth]{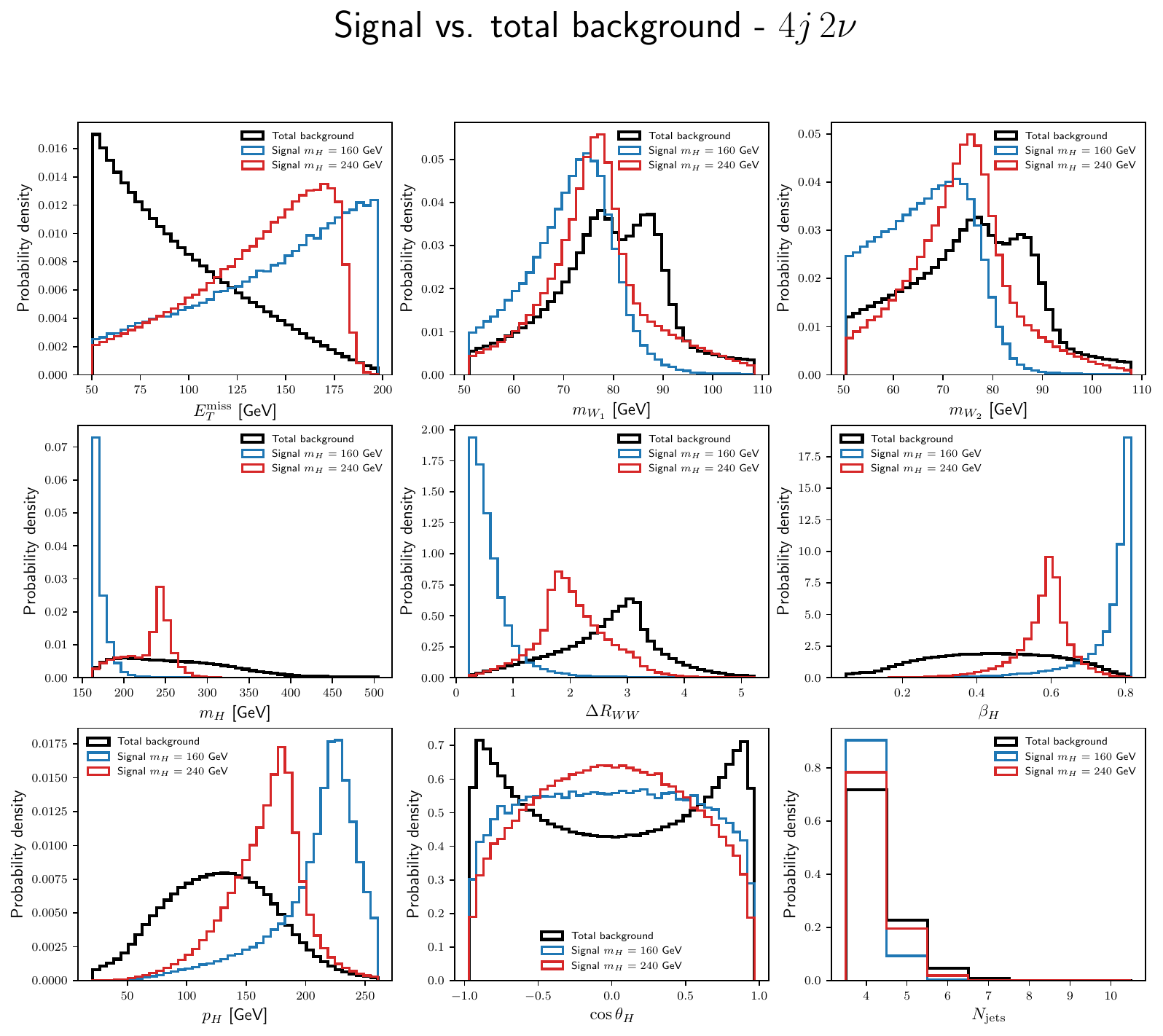}
		\caption{Unit-normalised distributions of the input features for
			signal ($\mH=160$ and $240~\mathrm{GeV}$) and total background, in
			the $4j2\ell$ (top) and $4j2\nu$ (bottom) channels.  The feature
			set is the one listed in Sec.~\ref{sec:selection}; the $4j2\nu$
			channel does not contain the lepton-specific and dilepton-mass
			variables.}
		\label{fig:matrix}
	\end{figure}

	Figure~\ref{fig:matrix} shows the unit-normalised distributions of
	the input features for signal and background in both channels.  The
	distributions motivate the choice of features in
	Sec.~\ref{sec:selection} and confirm that signal and background are
	separated by the same variables that the BDT selects.  The $4j2\ell$
	and $4j2\nu$ channels share most of their discriminating variables
	($m_{W_1}$, $m_{W_2}$, $\Delta R_{WW}$, $\beta_H$, $\cos\theta_H$),
	with the expected differences in the lepton-related and
	missing-energy observables.
	
	% =====================================================================

\clearpage
%\bibliographystyle{unsrtnat}
%\bibliography{BIB_TO_USE}

\end{document}